\documentclass[useAMS,usenatbib]{mn2e}
\usepackage{graphicx}
\usepackage{txfonts} 

\newcommand{\acap}{\noindent}

\newcommand{\msun}{\ensuremath{M_\odot}}
\newcommand{\lbsun}{\ensuremath{L_{\rm B,\odot}}}
\newcommand{\lksun}{\ensuremath{L_{\rm K,\odot}}}
\newcommand{\rsun}{\ensuremath{R_\odot}}

\newcommand{\Dmfif}{\ensuremath{\Delta m_{15}}}
\newcommand{\Mbmax}{\ensuremath{M_{\rm B,max}}}
\newcommand{\nira}{\ensuremath{^{56}{\rm Ni}}}

\newcommand{\kcc}{\ensuremath{k_{\rm CC}}}
\newcommand{\kaia}{\ensuremath{k_{\rm Ia}}}
\newcommand{\fia}{\ensuremath{f_{\rm Ia}}}
\newcommand{\nia}{\ensuremath{n_{\rm Ia}}}

\newcommand{\tage}{\ensuremath{\tau}}
\newcommand{\taun}{\ensuremath{\tau_{\rm n}}}
\newcommand{\tauni}{\ensuremath{\tau_{\rm n,0}}}
\newcommand{\taunx}{\ensuremath{\tau_{\rm n,x}}}
\newcommand{\taugi}{\ensuremath{\tau_{\rm GW,0}}}
\newcommand{\taugw}{\ensuremath{\tau_{\rm GW}}}

\newcommand{\tausf}{\ensuremath{\tau_{\rm SF}}}

\newcommand{\azero}{\ensuremath{A_0}}
\newcommand{\ace}{\ensuremath{\alpha_{\rm CE}}}
\newcommand{\aff}{\ensuremath{A}}
\newcommand{\msec}{\ensuremath{m_{2}}}
\newcommand{\mpri}{\ensuremath{m_{1}}}
\newcommand{\mwd}{\ensuremath{M_{\rm wd}}}
\newcommand{\mdd}{\ensuremath{M_{\rm DD}}}
\newcommand{\betaa}{\ensuremath{\beta_{\rm a}}}

\newcommand{\betag}{\ensuremath{\beta_{\rm g}}}
\newcommand{\lbk}{\ensuremath{\mathcal{L}_{B(K)}}}
\newcommand{\lk}{\ensuremath{L_{\rm K}}}
\newcommand{\lb}{\ensuremath{L_{\rm B}}}

\newcommand{\msf}{\ensuremath{M_{\rm SF}}}

\title[The rates of Type Ia Supernovae. II.] {The rates of Type Ia Supernovae. II. Diversity of events at low and high redshift.}

\author[Laura Greggio]{Laura Greggio$^{1}$\thanks{E-mail:
laura.greggio@oapd.inaf.it} \\
$^{1}$INAF - Osservatorio Astronomico di Padova, Vicolo dell'Osservatorio 5, I-35122 Padova}
\begin{document}

\date{Accepted ... Received ... ; in original form ...}

\pagerange{\pageref{firstpage}--\pageref{lastpage}} \pubyear{2002}

\maketitle

\label{firstpage}

\begin{abstract}
This paper investigates on the possible systematic difference of Supernovae Ia
(SN Ia) properties related to the age and masses of the progenitors that could
introduce a systematic bias between low and high redshift SN Ia's. 
The relation between the main 
features of the distribution of the delay times (DTD) 
and the masses of the progenitors is illustrated for the  
single (SD) and double degenerate (DD) models. Mixed models, which assume 
contributions from both the SD and DD channels, are also 
presented and tested versus the observed correlations between the 
SN Ia rates and the parent galaxy properties. 
It is shown that these correlations can be accounted for with 
both single channel and mixed models, and that the rate in S0 and E 
galaxies may effectively provide clues on the contribution of 
SD progenitors to late epoch explosions. A wide range of
masses for the CO WD at the start of accretion is expected in almost
all galaxy types; only in galaxies of the earliest types the properties
of the progenitors are expected to be more uniform. For mixed models, 
late type 
galaxies should host SD and DD explosions in comparable fractions, 
while in early type galaxies DD explosions should largely prevail. Events 
hosted by star forming galaxies span a wide range of delay times; 
\textit{prompt} events could dominate only in the presence of a strong 
star-burst. It is concluded that nearby SN Ia samples should well 
include the young, massive and hot progenitors that necessarily dominate
at high redshift.
%systems,
%and present various possible options within the close binary evolution 
%framework. I will also discuss the impa
%The distribution of the delay times (DTD) of SN Ia explosions plays a 
%fundamental role in determining the rate of events in stellar systems.
%In this paper I discuss the current literature propositions for the DTD, 
%focussing on the expectations from stellar evolution theory. Using an 
%analytical approach, I will illustrate how the main features of the DTD 
%are related to the distribution of masses and separations in the progenitor 
%systems,
%and present various possible options within the close binary evolution 
%framework. I will also discuss the impact of the different options on
%the observed rates.
\end{abstract}


\section[]{Introduction}
Type Ia Supernovae (SN Ia) provide an important contribution of iron to the 
interstellar medium \citep{mg86}. Because of this, the rate of SN Ia 
explosions, and
its secular evolution are fundamental for understanding abundances and
abundance patterns in stellar systems, in galaxy clusters and in the 
intergalactic medium. Due to the high energy release of each event, the rate of
SN Ia explosions also impacts on the evolution of gas flows in Elliptical 
galaxies \citep{cio91}, and
on the feedback process which follows from an episode of star formation (SF). 
The rate of nucleosynthetic and energetic input from SN Ia's critically 
depend on the distribution of the delay times (DTD), the delay time being
the time elapsed between the birth of the SN Ia progenitor star and its final
explosion. In the current literature, the shape of the DTD is still 
considerably debated (see \citealt*{grd} and references therein),
partly because the SN Ia progenitor's have not been unequivocally identified.

Most notably, Type Ia Supernovae are crucial cosmological probes, e.g. 
having provided the first evidence for cosmic acceleration \citep{riess,perl}. 
Their use in cosmology is
subject to the {\it standardisation} of their light curve, i.e. the ability 
to relate a distance independent quantity to the absolute magnitude at 
maximum light. For example, the Phillips relation links  
the SN Ia magnitude at maximum (e.g. \Mbmax) to its dimming 
in the first 15 days after maximum (\Dmfif, a measure of the decline 
rate) \citep{phillips,altav}.
While the relation very likely reflects different production of radioactive
\nira\ in the explosion \citep{arnett,mazzali}, 
the physical mechanism which causes this diversity 
is not clear, nor is the reason for the observed dispersion of  
the relation, part of which could be of intrinsic origin.
In most practical applications the {\it standardisation} of the SN Ia light 
curves involves the use of the {\it stretch} parameter ($s$,\citealt{perl}), 
which is related to the width of the light curve and the magnitude at
maximum light, based on observational data.    
Therefore, at present, the {\it standardisation} of the SN Ia light 
curves is fully 
empirical, and based on events at low redshift: if a systematic different 
relation holds for high $z$ events (either in the average value, or in the
dispersion, or both) the cosmological application of SN Ia's as distance 
indicators would be called into question.  Big, homogeneous samples of
nearby events are currently being constructed, from which robust correlations
can be derived, and peculiar events single out (e.g. \citealt{hicken}), but 
%identifying the origin of these correlations is crucial for the 
%cosmological application. 
ultimately, we want to assess 
whether the SN Ia light curve {\it stretching} works in the same way at
high and low redshift. It is then foremost important to understand 
the diversity of SN Ia events, and to check whether this is related to 
parameters like age and metallicity, which both increase systematically 
from high to low redshift.

Several observational attempts have been performed to clarify this issue. 
A relation between the \nira\ mass and the metallicity of the 
progenitor is expected  
on the basis of the nucleosynthetic processing \citep*{timmes};
this prediction has been studied by trying to determine the metallicity
of individual progenitors (e.g. \citealt{ellis}), as well as that of the 
stellar population hosts \citep{galla,howell}. So far the results are not 
conclusive,
and metallicity does not appear to be the most important parameter driving
the diversity (see \citealt{howell} for a thorough discussion). In this 
respect, it is worth noting that a sizable fraction of mass in stars in the
local universe belongs to populations that are both old and metal rich 
(\citealt{alvioara} and references therein); therefore also at high redshift
an important fraction of SN Ia events come from high metallicity stars.
%
%also at high redshift events will come from high metallicity
%progenitors, since elliptical galaxies are old and metal rich 
%
Thus, unless the metallicity dependence of the SN Ia light curve is very 
pronounced, the trend of its properties with redshift 
related to the change of (average) metalliticy should be weak.
More significant are the empirical correlations found between the light 
curves and the youth of the galaxy host. Literature results include the
average \Dmfif\ being higher in later galaxy types \citep{altav}; 
brightest events occurring preferentially (if not only) in hosts with high
SF rate per unit mass \citep{galla}; a correlation between the \nira\ mass
and the average age of the SN Ia host, with only faint events in galaxies
with average ages older than $\sim$ 3 Gyr \citep{howell,neill}. 
The observed trends all
point in the same direction, i.e. the younger the stellar population the more
energetic its SN Ia, but the details differ. For example, \citet{galla} find
that SN Ia's in galaxies with high specific SF span a narrow range in V 
magnitude at maximum, while the diversity is much more pronounced at low 
specific SF. On the other hand, \citet{howell} find a wide range of \nira\
mass in events hosted by young stellar populations, which suggests that
diversity is a feature of star forming galaxies. 
\citet{altav} find a wide range of SN Ia light curve properties in  
both early and late type galaxies; \citet{neill} find that faint 
events occur only in high mass galaxies, suggestive of a uniformity of 
SN Ia's in old stellar populations. Likely, the picture is confused because 
different authors use different proxies to characterise the light
curve of SN Ia's (stretch parameter, \Dmfif, magnitude at maximum in
some photometric band, ejected \nira), and the age of the parent 
stellar population (light or mass averaged age, spectral indices, morphological
type).   
The empirical route to study this issue is further 
complicated by the 
effect of extinction which is difficult to correct for in the individual 
events (e.g. \citealt{nobili}).

%Observational attempts to clarify this questions include the relation between
%the decline rate (as measured by the \textit{stretch } parameter $s$  
%\citep{perl} and the
%average age of the parent galaxy, as traced by morphology \citep{altav}, 
%or by the spectral energy distibution \citep{howell}. A systematic trend 
%between $s$ and the metallicity of the parent galaxy has also been 
%investigated
% \citep{galla,howell}.
%So far, the results indicate weak correlations, with bright events
%being hosted preferentially by relatively young, metal poor galaxies. However,
%different studies do not agree, the correlations (when existing) are weak, and
%depend on the tracers used for the decline rate and for the age or metallicity
%tracer (Galex, neill..). The empirical route to study this issue is further 
%complicated by the 
%effect of extinction which is difficult to correct for in the individual 
%events (e.g. \citealt{nobili}).

From the conjectural point of view, the claim of two 
populations for SN Ia's \citep*{mannu06}, one (\textit{prompt}) supplying 
events within the first 0.1 Gyr from the SF episode, and one (\textit{tardy})
providing explosions over the whole remaining age range (but with a much lower
efficiency) has suggested the existence of two distinct channels for SN Ia 
production, possibly characterised by different light curve parameters.
Since at high redshift the SN Ia production via the \textit{prompt} channel 
must prevail, a systematic evolution of the Phillips relation could be present,
related to the change of  mixture of \textit{prompt} and
\textit{tardy} explosions. 

From the stellar evolution point of view, 
while it is generally agreed that SN Ia explosions are due to the ignition of 
fuel in CO white dwarfs (WD), the specific  
path to the successful event is still debated. The lack of
hydrogen lines in most SN Ia spectra has traditionally lead to exclude single
stars (at the end of their evolution on the Asymptotic Giant branch) as 
viable candidates for SN Ia's. The explosion of a (naked) WD may be obtained 
via accretion from a close companion 
in a binary system, either as a consequence of Helium detonation, when the
accreted He layer reaches a critical mass (sub-Chandrassekhar models, 
hereinafter Sub-Chandra), or of Carbon deflagration, when the CO WD reaches
the Chandrassekhar mass (hereinafter Chandra) (see \citealt{hille}). 
Different progenitor models correspond to
different companions of the WD, which can be non degenerate (MS or post MS)
or degenerate stars (another WD); the relative evolutionary channels are 
commonly known as Single (SD) and Double Degenerate (DD), respectively. 
Therefore, stellar evolution in
a close binary system does provide two different channels for SN Ia production,
and also two different kinds of explosion. In each case, though, the
range of delay times can be very wide, from $\sim$ 40 Myr (the lifetime of an 
8 \msun\ star, i.e. $\sim$ the most massive CO WD progenitor) up to a Hubble 
time, so that each channel can accommodate 
\textit{prompt} and \textit{tardy} events. At the same time, a systematic
variation of the exploding WD parameters (e.g. the range of initial mass of 
the primary in successful systems) with the delay time is expected 
within each channel, and this could account for the observed diversity, as well
as imply correlations with the age of the parent stellar population.

From the observational point of view, there is no clear indication 
decisively supporting one channel over the other.
The circumstellar material detected for the event SN 2006X \citep{patat} 
has been interpreted as
favouring the Single Degenerate model, while the lack of it in other events
is more easily understood within the Double Degenerate model (see 
\citealt{simon}). According to \citet{kasen10} theoretical computations, 
the presence of a companion should cause an excess emission in the 
early light curve, which originates from the collision of the ejecta with 
the companion. The phenomenon, which should be detectable for $\sim$ 10 \% 
of the SD events, has not been reported so far, and provides a very 
interesting test of the progenitor model.
%Meanwhile, very luminous SN Ia events, like .... seem to require 
%Super-Chandrassekhar explosions, better explained with the merging
%of two WDs. 
Statistics of potential progenitors for the two models is also not conclusive. 
On the one hand there are many types of interacting binaries containing 
a WD plus a non degenerate star \citep{partha}, like cataclismic binaries, 
symbiotic systems and the particularly promising class of Supersoft X-ray 
sources, but it's not clear whether the population of good candidate SN Ia
precursors among them is abundant enough to account for the observed SN Ia rate
\citep{munari,kenyon,distefano}. On the other hand, the inventory of close 
double degenerates from the SPY project did yield one system with total mass
larger than the Chandrassekhar limit and close enough to merge within a Hubble
time \citep{napi}, showing that potential SN Ia progenitors can be
found among DDs as well. A few events have been reported to require a 
Super-Chandrassekhar WD mass \citep{howell06,yamana}: these are more easily 
explained as DD mergers. Meanwhile, single degenerate systems 
containing a WD with mass close to the Chandrassekhar limit have been 
discovered, like RS Oph \citep*{hachi} and 
HD 49798/RX J648.0-4418 \citep{mereghe}, which are very good candidate SN Ia 
precursors.  
These evidences may well suggest that both channels are  
providing SNIa events with comparable efficiency.

In this paper some general features of the DTD which are expected
from stellar evolution in close binaries are illustrated, following the 
analytical parametrisation presented in \citet{g05} (Paper I), and described in 
Sect. 2.
The dependence of the DTD on the major parameters is considered next, 
focusing on the possibility of realising
strong discontinuities which may justify the concept of \textit{prompt} 
and \textit{tardy} events. The systematic variation with the delay time of some
properties of the progenitors (i.e. WD cooling times, mass of the CO WD at the
start of accretion, total mass of the DD systems) is also examined. 
These two aspects
impact on the possibility of systematic differences between low and
high redshift samples of SN Ia, and are discussed in Sect. 3, separately for 
SD and DD models. Mixed (SD + DD) DTD models are presented in Sect. 4, 
where the possibility of constraining the mix from the 
observations is discussed. The results are summarised and discussed in Sect. 5;
conclusions are drawn in Sect. 6.  

\section[]{The analytic approach to the DTD}

The standard evolutionary scenario to SN Ia explosion starts with a close 
binary 
system in which the primary is an intermediate mass star, so that the first
Roche lobe overflow (RLOf) leaves a CO WD. Low mass stars 
($M \lesssim 2 \msun$) do produce CO WDs, but in a close
binary they can do so only if the first mass exchange takes place after the 
He flash. Since the stellar
radii at He flash are large (hundreds of solar radii), and since the 
distribution of the original separations $A_0$ scales as 1/\azero,
in most cases low mass primaries in close binary systems will fill their
Roche lobe (RL) while on the first Red Giant branch, thereby producing Helium, 
rather than CO, WDs. Therefore, reasonable limits to the primary mass (\mpri) 
in SN Ia progenitor systems are 2 and 8 \msun. When the secondary fills its 
RL there are two possibilities: either the CO WD accretes and burns the 
Hydrogen rich material remaining compact inside its RL, or a common 
envelope (CE) forms around the two stars. In the first case, the WD grows in 
mass and when conditions are met, a SN Ia explosion occurs; in the second case,
due to the action of a frictional drag force, the CE is 
eventually expelled, while the binary system shrinks \citep{it87}.
If the secondary is relatively 
massive, the outcome of the CE is a CO WD + Helium star system. 
Further evolution causes the He star to expand and fill its RL, pouring
Helium rich material on the companion, and the CO WD has again the possibility
of accreting and growing up to explosion conditions \citep{it94,wang}.
Both these paths to the SN Ia event are SD channels, but in the first one
(H-rich SD channel) some H is expected in the vicinity of the SN Ia which may 
show up in the spectrum, while in the second one (Helium star channel) no
H-rich material should be around.

If the CO WD does not accrete the donated Helium, a CE phase sets in again,
leading to a close CO+CO WD system. Angular momentum losses via gravitational 
wave 
radiation will eventually lead the system to merge, and, if its total mass 
exceeds the Chandrassekhar limit, explosion may arise. This is the
DD channel to SN Ia, as provided by intermediate mass binaries.
A DD system may also be formed when the secondary is a low mass
star, which fills its RL when its He core is degenerate. In this case, the
CE phase leaves a close binary with one CO and one He WD, but, typically, 
the total mass of the DD system does not reach the Chandrassekhar limit
\citep*{yungelson,belczy09}. Therefore this channel provides mainly 
sub-Chandra events; while potentially important for the SN Ia's, especially at 
late delay times, this path is neglected here because sub-Chandra explosions
do not provide a good fit to light curves and spectra of the majority of 
SN Ia's \citep{hoefli,nugent,hille}. 

Upon specifications of the distribution of the initial parameters (\mpri,\msec,
\azero), Binary Population Synthesis (BPS) codes compute the evolution of a 
population of binaries through these complicated paths, 
determine the partition of the population in the various channels,
and the delay times of the explosions.
The results are sensitive to the recipes used to follow the
evolution, most notably the criteria to decide what happens when a star
fills its RL. Critical prescriptions regard the occurrence of
CE or not, the outcome of mass exchange,
%(in both cases, whether CE occurs or not), 
the accretion modalities on the WD, whether fuel ignition leads to
accretion induced collapse or to a thermonuclear explosion \citep{ken85,
yoon,ken09,pier09}.
The DTD from simulations may present discontinuities due to the
switch on/off of one of the channels, and/or to an abrupt variation of the
efficiency with which one channel is providing SN Ia's, as a function of 
the original binary parameters. For example, a narrow DTD, centred on
a characteristic delay time, can be produced by suppressing 
all but one of the channels, and narrowing the range of the progenitor masses.
These (putative) discontinuities, and their location in delay time, 
follow from the prescriptions adopted in the BPS code. 
Many of the parameters which need
to be specified are very poorly constrained (e.g. the efficiency \ace\ of
the CE phase which controls the degree of shrinkage of the system); 
therefore the specific 
realisations can be \textit{up to date}, but remain uncertain.
%The straightforward way to compute the DTD is BPS.
%Many the potential channels for 
%SN Ia production in close binary evolution 
%are currently credited in the literature (refs..). These include
%SD H rich 
%channels, SD He* channels, DD channels, in many cases both Chandra and 
%Sub-chandra explosions. Now, with BPS codes one 
%AIC or SN Ia. Besides, other recepies need to be specified, including the
%computation of the RL, orbit synchronization, magnetic bracking, and 
%ultimately
%even stellar evolution (Mass loss, overshooting, initial-final mass relation).

\begin{figure}
\resizebox{\hsize}{!}{\includegraphics{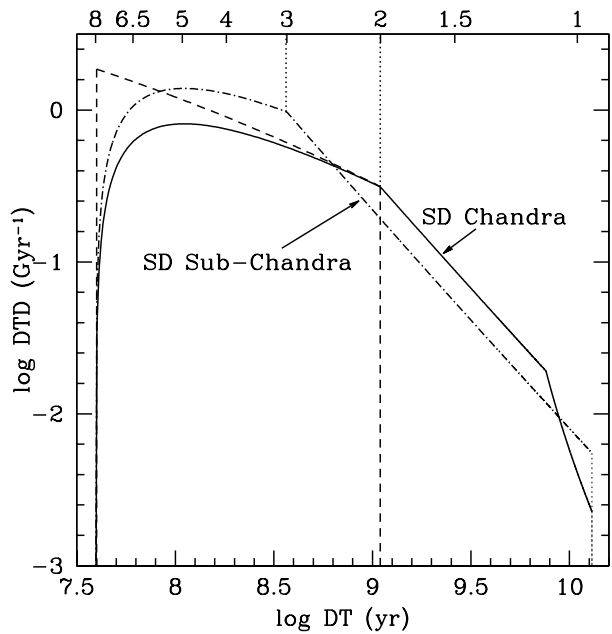}}
\caption{Distribution of the delay times for SD models,
Chandra (solid) and Sub-Chandra (dot-dashed). The DTD is
normalised to 1 over the range 0.04 to 13 Gyr. The upper axis is labelled 
with the value of the stellar mass (in \msun) evolving off the MS at the
corresponding a delay time labelled on the lower axis.
A Salpeter IMF for the primaries, a flat distribution of the mass ratios, and
$\mpri \geq$ 2 (3) \msun\ for Chandra (Sub-Chandra) models have been 
adopted.
The dashed line shows the trend of the CO WD formation rate from the 
primaries, 
which may represent an upper limit to the slope of the DTD 
for the MS+WD channel. This function, plotted on the logarithmic scale, 
is not normalised to 1 over the delay time range.}
\label{dtd_sd}
\end{figure}

An alternative approach to derive the DTD of SN Ia explosions consists in
characterising its shape using general stellar evolution arguments, and
introducing a parametrisation of a few astrophysical variables believed to 
play an important role. This approach was followed by \citet{gr83}
for the SD channel, improved in \citet{g96} to accommodate Sub-Chandra
explosions, and further developed in Paper I, where analytic
functions for the DTD are derived for both the SD and the DD channels.
Compared to the numerical results from BPS, these functions provide a more
flexible tool to investigate on the predictions from different progenitor 
models on various observables, thereby offering a way to constrain those 
astrophysical parameters which are recognised as crucial. On the other hand, 
in order to derive the analytic functions, some approximations and assumptions
have to be made, which are summarised in the following. We refer the reader
to Paper I for the detailed derivation of the analytic DTDs.

\subsection {The Single Degenerate channel}

Regardless the specific path, the delay time for SD channels is very close to 
the evolutionary lifetime of the secondary star off the MS (\taun(\msec)):
for the H-rich SD explosion the extra time spent to reach the RL dimensions, 
and the accretion timescale during which the CO WD grows to ignition 
conditions are both much smaller than \taun(\msec). For the 
Helium star channel, some correction (of about 10\%) should be
applied to account for the He burning lifetime of the former secondary; in
practise however \taun(\msec) fairly represents the total delay time also
in this case. Therefore, for the SD channel there is a one to one 
correspondence between the mass of the secondary and the delay time, so
that the distribution of the delay times is:

\begin{equation}
\fia(\tage) \propto n(\msec) \cdot |\dot{\msec}|
\label{eq_fiasd}
\end{equation}

\acap
where $n(\msec)$ is the distribution of the
secondary masses in systems which provide successful explosions through this
channel, and $|\dot{\msec}|$ is the rate of change of the stellar mass
evolving off the main sequence. The latter factor is a robust prediction 
of stellar evolution theory;
the first factor can be derived analytically by convolving the 
distribution of primary masses with the distribution of the mass ratios
($q = \msec/\mpri$) in binary systems. 
If all SN Ia's where arising from this evolutionary path, 
the range spanned by \msec\ should go from $\sim$ 8 \msun\ down 
to $\lesssim$ 1 \msun, to account for the wide range of delay times implied
by SN Ia events happening in all galaxy types \citep{gr83}.
The solid line in Fig. ~\ref{dtd_sd} shows the DTD for the SD Chandra channel 
under this hypothesis. The two cusps mark the 
setting in of two restrictions on the parameter space leading to explosion:
the first corresponds to the minimum mass of the primary, which is set to
2 \msun\ since the exploding WD should be composed of CO, rather than He
rich material. The later cusp is related to the requirement of totalling 
a Chandrassekar mass to
ensure explosion with a relatively low mass secondary: 
as the delay time grows long, the progressively lighter secondary
ought to combine with a progressively heavier CO WD (hence heavier \mpri)
to meet the Chandrassekhar limit. When the lower limit on \mpri\ 
becomes larger than 2 \msun\ the second cusp occurs. 
This restriction does not apply to 
Sub-Chandra events (dot-dashed line); thus, in this case, the DTD presents 
only one cusp, motivated by the requirement of a minimum
primary mass of 3 \msun\ as suggested by the models in \citet{stan94}.

A final remark concerns the WD + MS channel proposed by \citet{vandenh}: 
in this model
the companion fills its RL before core H exhaustion, and the delay time
could be appreciably shorter than the evolutionary lifetime of the secondary. 
The derivation of an analytic DTD is hardly possible in this case, 
since the time at which RLOf happens depends on the evolution of the binary
separation as determined by several physical processes, e.g. 
CE evolution, magnetic braking, tidal interactions \citep{iben91}. 
%whose effect can only be followed with simulations. 
However, the primary star still has to evolve and produce a CO WD; therefore,
for a given system, the delay time will be comprised between 
the evolutionary lifetimes of the primary and that of the secondary. 
If accretion starts soon after 
the formation of the CO WD, the distribution of the delay times will be given
by Eq.(\ref{eq_fiasd}) with \mpri\ replacing \msec, i.e. the WD 
formation rate
from the primary. If accretion is delayed until the secondary reaches core 
Hydrogen exhaustion the DTD will be given by 
Eq.(\ref{eq_fiasd}). The dashed line in Fig. \ref{dtd_sd} shows the CO WD 
formation rate (arbitrarily shifted) from primary stars with mass between
2 and 8 \msun, having assumed a Salpeter IMF. We may expect that the DTD for 
the MS+WD channel has an intermediate slope between the dashed and the solid 
line, i.e. more skewed at the short delays compared to Eq.(\ref{eq_fiasd}), but
not dramatically different.
 
%thus one can compute the distribution of the 
%\textit{minimum} delay times for SN Ia explosions via the MS+WD channel, 
%i.e. the distribution of the delay times for WD formation from the 
%primary. This is shown as a dashed line in Fig. \ref{dtd_sd} for primaries
%more massive than 2 \msun: compared to the solid line, this 
%distribution is more skewed at the short delays, but the slope is not
%dramatically different.   

\begin{figure*}
\resizebox{\hsize}{!}{\includegraphics{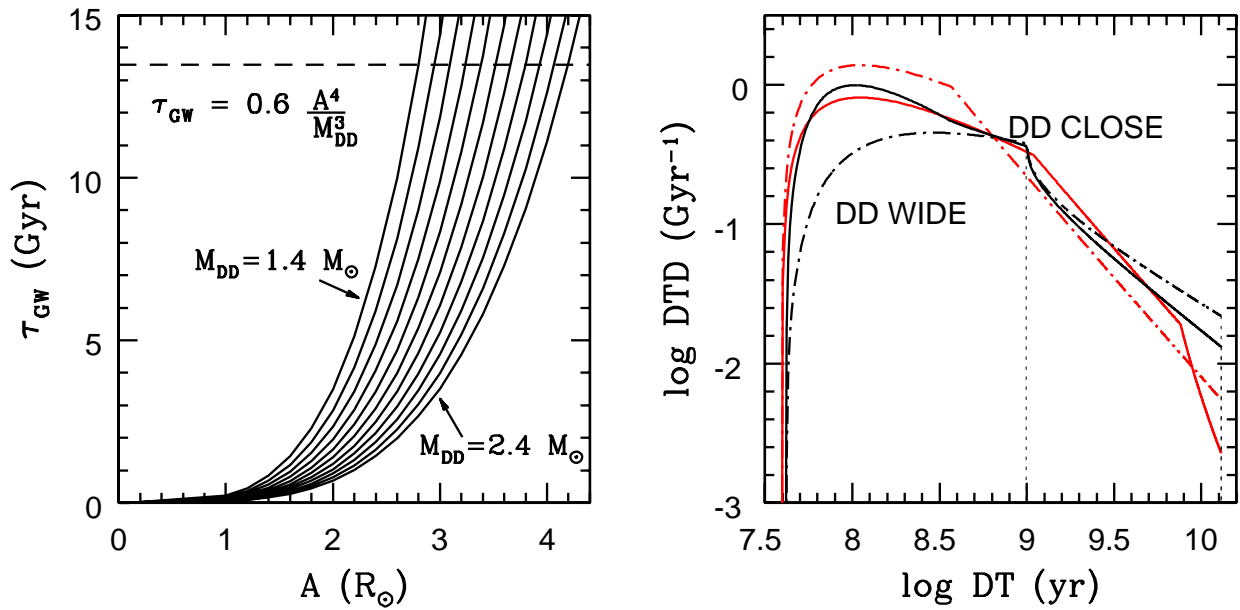}}
\caption{Left: the delay time due to the gravitational wave radiation, 
as given by the labelled (approximate) relation, as function of
the initial separation of the DD systems, and for 
\mdd\ ranging from 1.4 to 2.4 \msun\ in steps of 0.1 \msun.
Right: the distribution of the delay times for DD models,
CLOSE (solid) and WIDE (dot-dashed), in the same units as in Fig. ~\ref{dtd_sd}
. Adopted parameters include a Salpeter IMF for the primaries, 
a flat distribution of 
the mass ratios, $\taun \leq$ 1 Gyr, and flat distribution of the 
separations \aff.
The DTDs for the SD models are plotted in red for comparison.} 
\label{dtd_dd}
\end{figure*}

\subsection {The Double Degenerate channel}

The delay time for the DD systems is the sum of the MS lifetime of the
secondary (\taun(\msec), hereafter \taun) plus the time taken by the 
gravitational wave radiation to
get the DD system into contact (\taugw). The latter time delay depends on
the separation (\aff) and masses of the components of the DD system at birth;
if both \mpri\ and \msec\ are intermediate mass stars ($\geq 2 \msun$)
this delay can be approximated as 
$\taugw = 0.6 \, {\aff^4} \, {\mdd^{-3}} \, \mathrm{Gyr}$,
where \mdd\ is the total mass of the DD system, and both \aff\ and \mdd\ are 
in solar units (Paper I). For what explained above, 
this limit on the masses of both binary components corresponds to considering
CO + CO DD systems; besides, the limit \msec $\geq$ 2 \msun\ 
implies a maximum delay \taun\ of 1 Gyr. 
The delay \taugw, plotted in the left panel of Fig. ~\ref{dtd_dd}, 
is very sensitive to 
both \aff\ and \mdd, and in principle it ranges from virtually zero, 
for the most massive / closest systems, up
to infinity for the least massive / widest ones. Fixed limits apply to
\mdd, from 1.4\msun\ (Chandra explosions) to $\sim 2.4 \msun$, when both 
components in the primordial system are of 8 \msun \footnote{The remnant of
an 8 \msun\ star is assumed to be a CO WD of 1.2 \msun, consistent with
the empirical relation in \citet*{willi} and in \citealt{kalirai}}. 
More elusive is the allowed range for \aff, although it must
be smaller than $\sim 4 \rsun$ in order to provide events within a Hubble time.

Typically, the separation of the DD systems at birth is the result of 
evolution through at least 2 RLOf
episodes. Given the difficulty of describing this complex hydrodynamical
phase, the results of the application of the various recipes in the literature
are uncertain. In this respect, it is important to notice that a  substantial
shrinkage is necessary to account for SN Ia events within a Hubble time
from DD systems, since the first RLOf occurs when the separation \azero\
ranges from several 10 \rsun\ (to avoid premature merging), 
to several 100 \rsun\ (to get RLOf at all, see \citealt{it87}).
Therefore, the prescription used in BPS codes to determine the 
degree of shrinkage should be accurate at a level of $\sim$ 1 \%  
to yield correct results on an individual binary.

The sensitivity of the ratio $\aff/\azero$ to the commonly adopted literature
descriptions of the CE phase is explored in Paper I. It appears that when 
the \ace\ formalism \citep{webbink} is applied to two mass exchange phases, 
the ratio is very small, so that only initially wide systems manage to populate
the relevant range of separations (up to a few \rsun). Conversely, if the 
first of the two CE phases is regulated by the \citet{nele00} 
(angular momentum loss) recipe, 
%which leads to a less efficient shrinking,
the ratio $\aff/\azero$ covers a $\sim$ 2 orders of magnitude range, depending
on the masses of the components.
Moreover, in the former case, the heavier the original binary the lower
the $\aff/\azero$ ratio: this is because in the \ace\ formalism
the shrinkage at each CE phase is proportional to the binding
energy of the envelope of the donor, which is larger when the donor is more
massive. Thus, the larger \msec, the shorter \taun\ \textit{and} the shorter
\taugw: this skews the DTD towards the short delay times.
When the Nelemans' formalism is applied, instead,  
no strong relation appears to emerge between the initial masses of the binary
components and the shrinkage, so that \taun\ and \taugw\ are likely to be
uncorrelated, as well as \mdd\ and \aff.

On this premises, analytic distributions of the delay times for the DD channel
have been derived in Paper I in two flavours: the DD-CLOSE model, in which
at each \msec\ the delay time varies between a minimum (=\taun(\msec)) and a
maximum delay, whose value increases as \msec\ decreases. 
In this scenario, a wide
portion of the initial parameter space (\mpri,\msec,\azero) maps into short
delay times, while the long delay times are populated with only the less
massive binaries. The DD-WIDE model, instead, assumes that \aff\ and \mdd\ are
independent variables, and their distributions concur to shape the distribution
of the gravitational wave radiation delay. 
The variables \taun\ and \taugw\ are
also assumed to be basically independent, except for a correlation between
\msec\ (i.e. \taun) and \mdd, to account for the fact that the heavier the
original secondary, the more massive the DD remnant. As a result, in
the DD-WIDE channel the upper limit to \taugw\ is greater then the Hubble time
irrespective of the mass of the original binary, and the population at long
delay times of the corresponding DTD is enhanced. Simple power laws are 
assumed to describe the distribution of \taugw\ (CLOSE DDs) 
and \aff\ (WIDE DDs), with parametrised exponents respectively called 
\betag\ and \betaa. 
The shape of the distribution of \mdd\ is found to be 
relatively unimportant.

Admittedly, the above approximations and the assumptions are rough and do not 
punctually describe the 
outcome of the evolution of close binaries, SN Ia progenitors. However, 
the range spanned by \mdd\ and \aff\ for events within a Hubble time are small,
so that the assumption of smooth
distributions of these parameters is likely appropriate.
The main features of the curves
are determined by the gravitational wave radiation clock, which implies that
the distribution of \taugw\ is skewed at the short end, unless all successful 
DDs are born very wide. This characteristic can be
judged by projecting
a flat distribution of \aff\ into the corresponding distribution on \taugw\
in Fig. ~\ref{dtd_dd}, left panel. 

The right panel of Fig. ~\ref{dtd_dd} shows the DTD for CLOSE DDs and 
WIDE DDs, normalised to 1 over the 
range of delay times up to 13 Gyr. The distributions can be
viewed as a modification of the SD case, where systems with given \taun\ 
populate a range of delay times longer than this due to the extra delay
\taugw. The WIDE DD case is the least populated at the short delay times,
the CLOSE DD the most, since in this case the great majority of systems 
have a short \taugw. 
Notice that delay times in excess of 1 Gyr are populated only with systems 
whose
\taugw\ is longer than some limit, since by construction a maximum \taun\ of
1 Gyr is imposed. This is the reason for the presence of the cusp in the
DTDs: at this delay time a fixed limit on the allowed parameter (\taun) space 
is hit.
Although this feature in the analytic DTDs is purely numerical, we do
expect an abrupt change in the distribution at a delay time equal to the 
MS lifetime of the least massive secondary in successful 
SN Ia progenitors.  
  
Inspection of Fig. \ref{dtd_dd} shows that the various models provide similar
curves: the DTD features an early maximum and most events occur within 
$\sim$ 1 Gyr from formation. At later epochs the DTD exhibits a steady decline
with a pronounced downturn for the SD Chandra model. 
All distributions accommodate \textit{prompt} and \textit{tardy} events,
but with a substantial continuity over the whole range of delay times;
this follows from the assumption of continuous distributions over the relevant
parameter space of masses and separations of SN Ia progenitors.
In the next section these curves are compared with theoretical results in the 
literature.

\subsection {Analytic DTDs and BPS results}

In deriving the analytic DTDs described above one potentially important 
variable is neglected, i.e. the separation of the primordial binary \azero.
In order to provide a successful event, \azero\ ought to range between
a minimum and a maximum value, respectively to avoid premature merging, and 
secure interaction at all. As long as these limits do not depend on the
component's masses, the shape of the DTD is not affected, and the limitation
just reflects into the overall likelihood of the SN Ia
event from a stellar population. However, the minimum and maximum initial 
separations tend to increase with the mass of the 
primary, because more massive stars have larger radii at all key points
along evolution. Figures 1 to 4 in Iben and Tutukov (1987) well illustrate
this point, also showing that, in general, the range of separations for a 
successful interaction widens as the primary mass grows. It the  
distribution of \azero\ were flat, the more massive binaries would have then 
greater chance to provide SN Ia's; on the other hand, closer systems should 
be more abundant \citep{abt}, so that the more massive binaries also have a 
higher likelihood of premature merging.
Therefore, it is not clear how the efficiency of SN Ia production from a 
binary of given (\mpri,\msec) varies with the system separation.
The comparison of the analytic DTDs with results from BPS
helps understanding whether or not the analytic functions capture the most 
important astrophysical parameters.

In Paper I the analytic DTDs are compared to the BPS results of 
\citet{yungli}: consistency is found for the Sub-Chandra and for 
the DD-Chandra cases, when accounting for the limits on 
\mpri\ and \msec\ adopted by the authors. The comparison between the DTDs of
the SD-Chandra channel is hampered by the narrowness of the range of 
secondary masses in systems which provide SN Ia through the H-rich SD path 
in the \citet{yungli} simulations. Actually, the conditions for stable mass
transfer and/or a successful growth to the Chandrassekhar limit of the 
H-accreting CO WD 
are often found to hold only within a very restricted range of \msec.
The analytic DTDs do not incorporate constraints on the mass transfer rate,
but the mass range covered by the components of successful systems is to be
considered as a parameter.

\citet*{hkn08} present the DTD for the SD-Chandra scenario taking into account
the effect of a radiative wind from the accreting WD which stabilises the mass
transfer \citep*{hkn96}, favouring the WD mass growth. 
Although their computation is not a BPS simulation, the DTD
is obtained by integrating the distributions over the appropriate range in the
parameter space (\mpri, \msec, \azero) which leads to successful explosions, 
according to
their modelling of the evolution of CO WDs accreting from a MS and from a 
RG companion. It is instructive to compare their derived DTD with the Paper I
analytical distributions.
%\citet{hkn08} present the DTD for the SD scenario taking
%into account the limits of the binary parameters  which
%result from their computation of the fate of a CO WD which accreting from a 
%MS and a RG companion, including the effect of a radiative wind from the WD.
Let's first consider the WD+RG channel: \citet{hkn08} find it active for 
\msec $\lesssim$ 3 \msun, and indeed the corresponding DTD is populated at 
\tage $\gtrsim$ 0.4 Gyr; its DTD is marked by two cusps, at 1.6 and 12.6 Gyr. 
Given the wide bins of delay time in the \citet{hkn08} 
computation, the position of the cusps are consistent with those on the
solid line in Fig. ~\ref{dtd_sd}. For the same choice of
IMF (Salpeter) and distribution of mass ratios (flat) as in Fig. ~\ref{dtd_sd},
the slope of the DTD between the two cusps (in logarithmic units) is -1.4, 
perfectly consistent with the analytic DTD. The WD+MS channel is found active 
for 2 $\lesssim$ \msec/\msun $\lesssim$ 6 and, in fact,
the relative DTD goes to zero in the vicinity of \tage = 1 Gyr; the slope is
$\sim$ -1.4, much steeper than the slope of the WD formation rate (dashed line
in Fig. ~\ref{dtd_sd}) in the
same range of delay times, which is $\sim$ -0.5. This may signal that the  
distribution of the separations is important in shaping the DTD of the 
WD+MS channel, favouring SN Ia events in the more massive binaries.
 
The DTD from the WD+MS channel has also been presented by \citet{han04}, as
derived with their BPS code. The resulting DTD is a
rather narrow, almost flat distribution, within limits which 
depend on the parameter for CE efficiency. A similar shape is found for the
WD+MS channel in \citet*{meng}. The limits are consistent with the MS
lifetime of the most and least massive secondary in SN Ia progenitors through
this channel, which are found confined between $\sim$ 3 and $\sim$ 2 \msun. 
These DTDs are quite different from those in \citet{hkn08}, possibly because 
of the use of a different technique and of different details in the 
assumptions. Therefore, the shape of the DTD for the WD+MS channel seems 
rather sensitive to the adopted approach.  
%The shape of
%the DTD for this channel appears rather sensitive to the various assumptions
%necessary to trace individual binaries down to explosion.

%DTD based on BPS calculations have also been presented by Belczynski, 
%Bulik and Ruiter 2005 for three SN Ia channels: a single degenerate which
%includes both Chandra and Sub-Chandra explosions (SDS); and two double
%degenerate channels, one populated with Sub-Chandra (SWB) and one with Chandra
%(DDS) explosions. The DTD of the latter channel is found proportional to
%$\tage^{-1}$ (for an efficient shrinkage during the CE phase), slighlty 
%flatter ($\propto \tage^{-0.8}$).

DTDs based on BPS calculations have been presented by \citet*{belczy05} and 
more recently by \citet{belczy09}. Since the former models include Sub-Chandra 
explosions, which actually provided the majority of events 
\citep{belczy09}, their comparison 
with the analytic DTDs is not practical. More similar are the SN Ia 
production channels in the models presented in \citet{belczy09}, which include
only Chandra exploders, from DD progenitors (DDS), H-rich SD Chandra 
progenitors (SDS), and CO WDs accreting Helium via stable RLOf,
either from a non degenerate or from a degenerate companion (AM CVn). 
The DTD for the DDS channel is very similar to the analytic function, with a
logarithmic slope of $\simeq$-1. Notice that when the efficiency of the CE 
phase is reduced (\ace=0.5) the DTD of the DDS channel is found steeper, as
a consequence of the smaller orbital separations of the DD systems when they
emerge from the CE. The same effect can be obtained by varying the parameter
\betag\ or \betaa\ in the analytic DTDs. 
The comparison with the BPS results for the SD
channel is more complicated, due to its splitting into the two classes SDS
and AM CVn in \citet{belczy09}. However, the BPS DTD seems much flatter than
the analytic DTD in the range of delay times between 3 and 6 Gyr, and lacks the
dramatic drop at late epochs. Since the clock of the explosion is the 
evolutionary lifetime of the secondary (also in the BPS simulations), it seems
that the envelope mass of the secondary has little influence on the efficiency
of the production of Chandra explosions in the SDS channel, in the simulations
by \citet{belczy09}.
 
To summarise, the comparison of the analytic curves with other results in the
literature shows that the limits on the component's masses play a crucial role 
in shaping the DTD, while the effects of the limits on \azero\ are much less 
evident, except perhaps for the WD+MS progenitors.
While the analytic DTDs for the DD models match well the BPS results,
the DTD of SD models turns out appreciably different. It seems however that
diverse realisations are obtained by different authors, both for the range
of \msec\ of SN Ia progenitors, and for the DTD slope. These differences
likely reflect the sensitivity of the results of the systems' evolution to 
details in the prescriptions adopted for the simulations.

%All in all, it doesn't appear that the separation plays a crucial role in 
%shaping the DTD within a given channel, except perhaps for the WD+MS p
%rogenitors.

\begin{figure*}
\resizebox{\hsize}{!}{\includegraphics{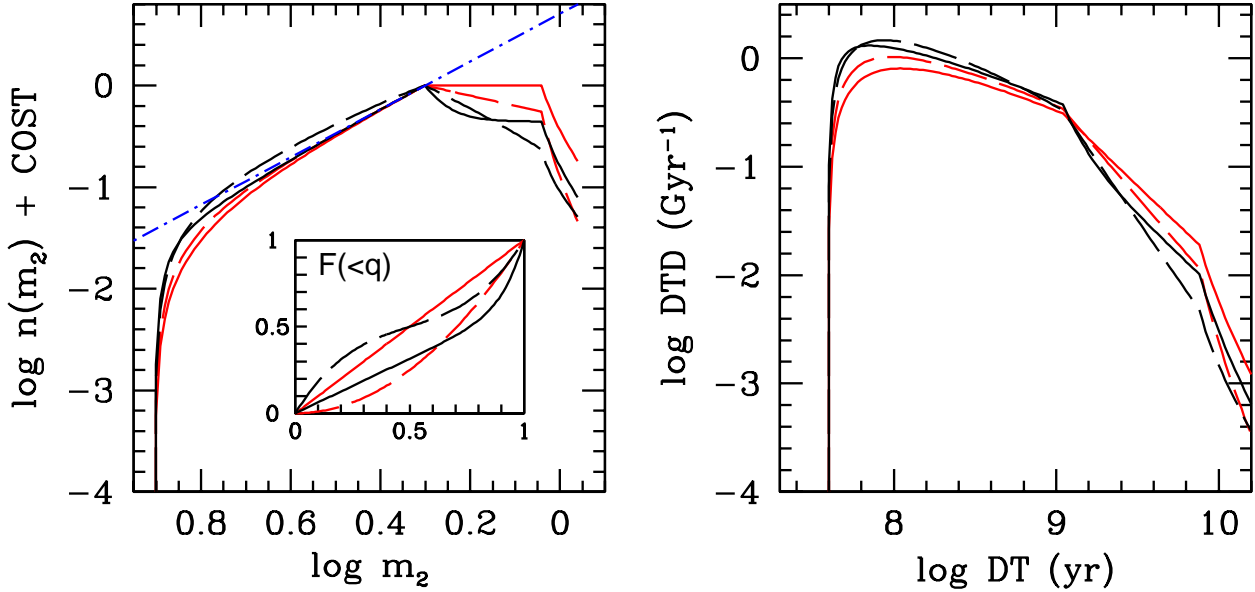}}
\caption{Effect of the distribution of the mass ratios on the mass distribution of the secondaries 
(left) and on the DTD (right) of SN Ia from the SD-Chandra channel. 
All curves assume (\mpri,\msec) $ \leq 8 \msun$, and a Salpeter
distribution for the primary masses. Red lines assume 
$f(q) \propto q^\gamma$ with $\gamma$ = 0 (solid) and 1
(dashed). Black lines assume Eq.(\ref{eq_fpi}) (solid) and
(\ref{eq_ftri}) (dashed). The cumulative distributions of $q$ 
are shown in the inset (left panel). 
The dash-dotted blue line in the left panel shows a Salpeter
distribution: with respect to this line, $n(\msec)$ is underpopulated
at both the high and the low mass end, because of the requirements on
the binary system SNIa progenitor. The deficiency at the high mass end is
due to the upper limit of 8 \msun\ on the primary mass; at the low mass
end, instead, it is due to the lower limit on the mass of the WD.}
\label{sd_parama}
\end{figure*}

\begin{figure*}
\resizebox{\hsize}{!}{\includegraphics{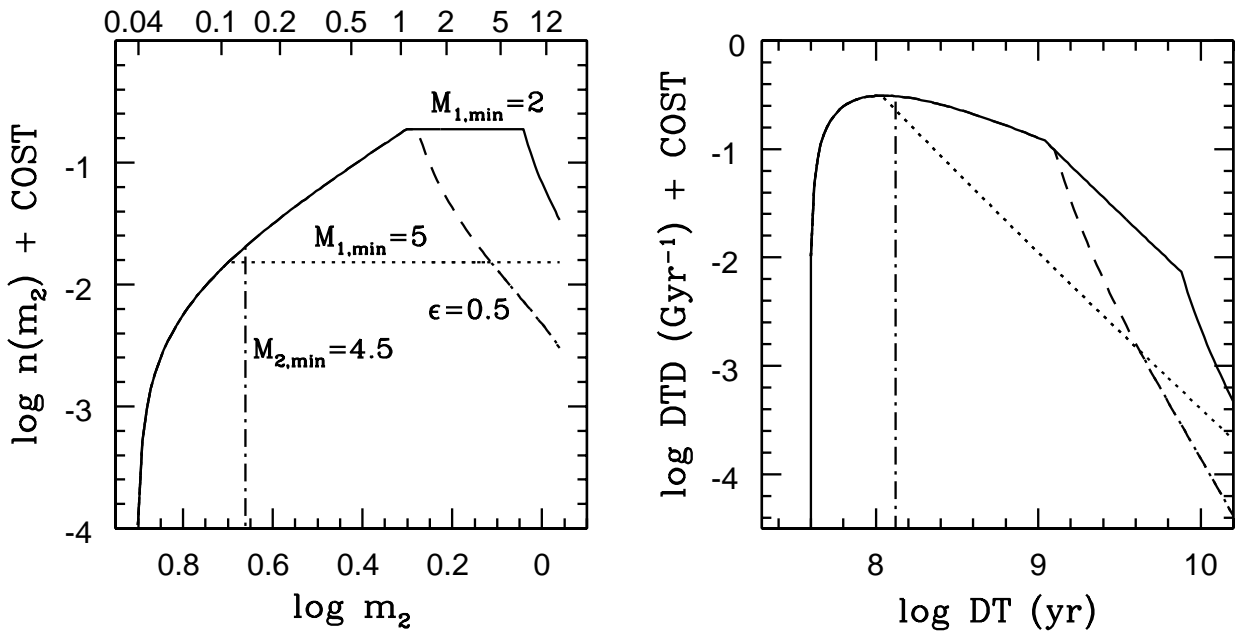}}
\caption{Distribution of the secondaries 
(left) and on the DTD (right) of SN Ia from the SD-Chandra channel for
a flat $f(q)$. 
All curves assume (\mpri,\msec) $ \leq 8 \msun$, but differ in 
%adopted minimum \mpri\ or \msec\ in successful systems, and  
%the efficiency of accretion on the CO WD ($\epsilon$). 
the value of other parameters: the solid and dotted
lines assume a minimum \mpri\ of 2 and 5 \msun\ respectively, no lower
limit on \msec, and $\epsilon=1$; the dashed line assumes the same limits as
the solid line, but $\epsilon=0.5$; the dot dashed lines assumes a minimum
\msec\ of 4.5 \msun. The line encoding is the same in the two panels. 
The top axis in the
left panel is labelled with the MS lifetime of the evolving secondary, in Gyr.}
\label{sd_param}
\end{figure*}

\section[]{Effect of parameters}

By construction, the analytic DTDs have a built in parametrisation of some
astrophysical variables related to the binary population and to the 
progenitor's model. In addition, they are constructed by relating the delay 
time to the exploding system, so that systematic variations of the properties 
of SN Ia progenitors with the delay time can be envisaged. 
In this section these issues are illustrated for the Single and the Double
Degenerates.

\subsection[]{Single Degenerates}
  
The curves in Fig.~\ref{dtd_sd} are constructed upon specification of:
 the slope of the mass distribution of the primaries 
($\alpha$), the distribution of the mass ratios ($f(q)$), 
the mass range of the primaries 
and of the secondaries in successful systems. An additional 
parameter is the accretion efficiency $\epsilon$, namely the fraction of the
envelope mass of the secondary which is actually accreted by the primary.
The solid curve in Fig.~\ref{dtd_sd} assumes $\alpha=2.35$ 
(as all other DTDs in this paper), a flat 
distribution of the mass ratios, the maximum possible range 
for \mpri\ and \msec, and that the whole envelope of the secondary is 
available for the WD to grow up to the 
Chandrassekhar limit ($\epsilon=1$). The same curve is reproduced as a
solid line in Fig.~\ref{sd_param}.
The dependence on $\alpha$ 
and $f(q)$ have been illustrated in Paper I: basically, any combination 
which favours low mass secondaries leads to a larger fraction of late epoch 
events. In Paper I it is also shown that when using the alternative algorithm 
adopted in \citet{gr83} to derive $n(\msec)$, the
fraction of low mass secondaries is enhanced, for a given IMF and 
distribution of the mass ratios \footnote{Notice that the \citet{gr83} 
formalism 
assumes power law distributions for the total binary mass $M_{\rm B}$ and 
for the ratio 
of \msec\ to $M_{\rm B}$, so that the same IMF slope and distribution of
the mass ratios have different meanings in the two approaches.}.

Empirical deteminations of the distribution of the mass ratios
in binaries are subject to considerable observational biases (see e.g. 
\citealt{kouwen}).
Most binary population synthesis computations assume a flat
distribution; on the other hand, there is growing evidence for 
a prominent population of twins among close binaries, which could
enhance the fraction of systems with short delay times
(\citet{pinso} and references therein).  
Fig. ~\ref{sd_parama} shows the distribution of the secondary
masses and the ensuing DTDs under various assumptions for
$f(q)$.  In particular, the two black lines assume a distribution of
the mass ratios made of two components: the solid line adopts

\begin{equation}
f(q) = 0.63 + 2.42 \times (q/0.95)^{10}
\label{eq_fpi}
\end{equation}

\acap
which is a flat distribution up to $q \simeq 0.8$, plus a prominent
population of twins. This function roughly fits the distribution in 
\citet{pinso} for $q \geq 0.5$. The dashed black line
adopts

\begin{equation}
f(q) = 6.9 q^2 - 6.9 q + 2.15
\label{eq_ftri}
\end{equation}

\acap
which roughly reproduces \citet{trimble}'s results, with a strong
component at the lowest and highest values of the mass ratio, and a
minimum at $q=0.5$. Clearly, the distributions with a high proportion
of twins (the two dashed lines and the solid black line) provide DTDs 
which are relatively more populated at the short
delay times. The effect however does not appear dramatic, nor does a
two components $f(q)$, as in Eqs. (\ref{eq_fpi}) and (\ref{eq_ftri}), 
reflect into a two components DTD.
All following computations adopt a flat distribution of the mass
ratios, an option which maximises the fraction of late epoch
explosions. This will be taken into account when discussing the 
results.

Similar to Fig.~\ref{sd_parama}, Fig.~\ref{sd_param} illustrates 
the effect of varying the other parameters.
%The solid curves in Fig. \ref{sd_param} assume the maximum possible range 
%for \mpri\ and \msec, and that the whole envelope of the secondary is 
%available for the WD to grow up to the 
%Chandrassekhar limit; 
The dashed curves assume the same limits for the masses 
of the components as the solid curve, but $\epsilon=0.5$ for all systems. 
The possibility of 
accreting only half of the envelope of the secondary has a dramatic impact 
on the number of systems with a low mass secondary which manage a Chandra 
explosion; correspondingly, the DTD drops rapidly off at delay times
slightly over a Gyr. This makes it difficult for the SD channel to account for
a sustained SN Ia rate at long delay times: for example, at an age of 13 Gyr, 
the stars evolving off the MS for a single burst stellar population have a 
mass of $\sim 0.9 \msun$. With a core mass of 0.3 \msun, only 0.6 \msun\ are
available for accretion onto the companion WD, and, if $\epsilon=0.5$, 
the Chandrassekar limit can be met only starting with a CO WD heavier than
1.1 \msun. Therefore, the range of suitable primary masses is exceedingly small
($7 \lesssim \mpri \lesssim 8$).
The dotted curves in Fig. \ref{sd_param} show the effect of implementing a
more severe lower limit to \mpri: 5 \msun\ instead of 2 \msun, corresponding 
to SN Ia's arising only from CO WDs more massive than $\sim$ 0.9 \msun.
In this case, $n(\msec)$ gets substantially deprived of systems as soon as
$\msec$ drops below 5 \msun, and the DTD gets depressed at the corresponding
MS lifetime ($\simeq$ 0.1 Gyr); however the distribution remains wide, without
strong discontinuities. A strong discontinuity may instead be obtained by
implementing a dramatic change in SN Ia production at some 
value of \msec, due to the one to one correspondence between delay time
and mass of the secondary in the SD model.
In particular, a lower limit on \msec\ has
the effect of vanishing the DTD at delays longer than the MS lifetime of such 
stellar mass.%(e.g. the dash-dotted lines in Fig.~\ref{sd_param}): i
 The dashed-dotted lines in Fig. ~\ref{sd_param}  
show the effect of adopting a lower limit of 4.5 \msun\ to the secondaries
in SN Ia progenitors, mimicking what happens for the SD Helium star channel
which requires massive secondaries. Notice that, according to \citet{wang},
the secondary mass in such SN Ia progenitors ought to be larger than 
5.6 \msun, implying a drop off in the DTD at $\sim$ 80 Myr. 

%This is what often happens in BPS realizations, where,
%when the donor is a low mass star, the mass transfer rate is found too high 
% for
%the WD to remain compact and a CE forms, aborting the WD growth to ignition
%conditions (ref...non l'ho mica trovata; mi sa che tocca guardare Nomoto 
%stesso....).
I turn now to consider some properties of the exploding WD which are predicted 
to vary systematically with the delay time in the context of the SD model.
% asillustrated in 
%Fig. \ref{sd_syst} shows how the range of the initial mass of the CO WD in
%varies as a function of the delay time.
\citet{lesaffre} 
presents a systematic study of the sensitivity of ignition conditions for 
H-rich Chandra SD exploders on various properties of the progenitor.  
These authors find that the more massive and/or the cooler the CO WD when
accretion starts, the higher the central density
at ignition. While it is not clear whether this leads to a larger or lower 
mass of synthesised \nira, a systematic variation of ignition density 
could translate into a systematic variation of the properties of the SN Ia
explosions. It is thus interesting to analyse how the initial mass of the 
CO WD and its cooling time vary with the delay time.  

\begin{figure}
\resizebox{\hsize}{!}{\includegraphics{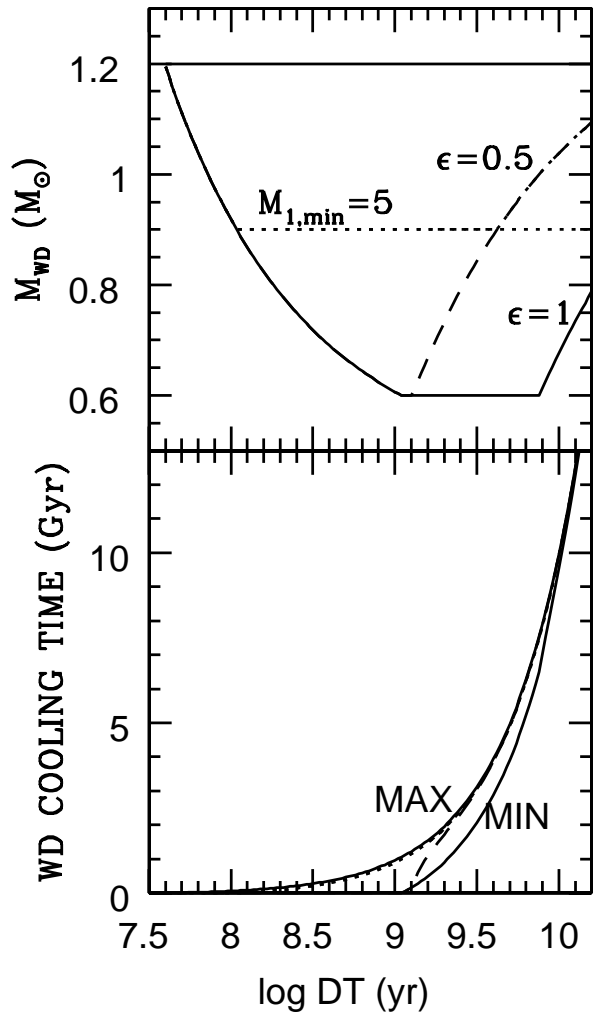}}
\caption{SD model: systematic variation with the delay time of the mass 
range (top) and of the cooling time range (bottom) of the CO WD when 
accretion starts. The different curves refer to different choices of the
minimum \mpri\ and of the parameter $\epsilon$ with the same line
encoding as in Fig. \ref{sd_param}.}
\label{sd_syst}
\end{figure}

The top panel in Fig. \ref{sd_syst} shows the range for the initial 
WD mass of SN Ia progenitors as function of the
delay time, for a few choices of the relevant parameters. The limits on
the WD mass follow from the limits on \mpri\ through the adopted 
initial-final mass relation ($\mwd=0.6+ 0.1 \times (M-2.)$ as in Paper I).
The maximum value of the primary mass is 8 \msun\, irrespective of the delay
time; the lower limit is given by \msec\ (function of \tage), or higher, 
either because a fixed limit on \mpri\ sets in (2 \msun\ for the solid line,
5 \msun\ for the dotted line), or, at late delay times, 
because of the need of securing a Chandrasekhar mass at explosion.
Correspondingly, the range of initial mass for the CO WD is small 
at short delays (only massive binaries have had time to evolve), but it 
rapidly increases with \tage; at intermediate delay times 
the range of initial WD mass is maximum, but at late delays it 
shrinks, because a massive primary is required to accomplish a Chandra 
explosion. 
The latter regime sets in earlier for a less efficient accretion 
(e.g. the dashed curve).

The bottom panel of Fig. \ref{sd_syst} shows the range of cooling times of
the CO WD, i.e. the time elapsed between its the formation and the start 
of the accretion phase. In this time interval the WD cools,
becomes progressively more degenerate, and other processes like C and O 
separation and crystallisation (e.g. \citealt{salaris}) set in. 
The path towards explosion, and the
explosion outcome critically depend on the initial structure of the CO WD,
or, in turn, on the cooling time. For example, the C and O relative diffusion 
may affect the amount of synthesised \nira, hence the light curve.
On average, the cooling time increases
with the delay time, and its range reaches a maximum at \tage $\simeq$ 1 Gyr,
to become smaller at longer \tage. Most noticeably, the cooling time is
very long, up to several Gyr, for most values of delay time.

To summarise, some dispersion in both parameters is present at
all delay times, but for early events this dispersion is more pronounced.
Qualitatively, then, we may expect more diversity in SN Ia events from younger
stellar populations. On the other hand, at short delays the CO WD at the 
beginning of the accretion phase is (on average) more massive and younger; 
thus, due to 
compensating effects, the trend of ignition density with delay time could be 
relatively weak. Conversely, at long delay times the mass and cooling age of
the CO WD both increase with \tage, so that the ignition density should 
increase systematically.
It is important to notice that very long cooling times, 
of few to several Gyrs, characterise most of the events, whereas
\citet{lesaffre} models include cooling ages of $\simeq$ 1 Gyr at most. 
As the delay time
increases, the CO WD has time to become more and more degenerate before the
start of the accretion phase. The evolution of the WD structure during the 
accretion phase, as well as the outcome of C ignition under extremely 
degenerate conditions are important questions which need more attention than 
currently present in the literature.

\begin{figure}
\resizebox{\hsize}{!}{\includegraphics{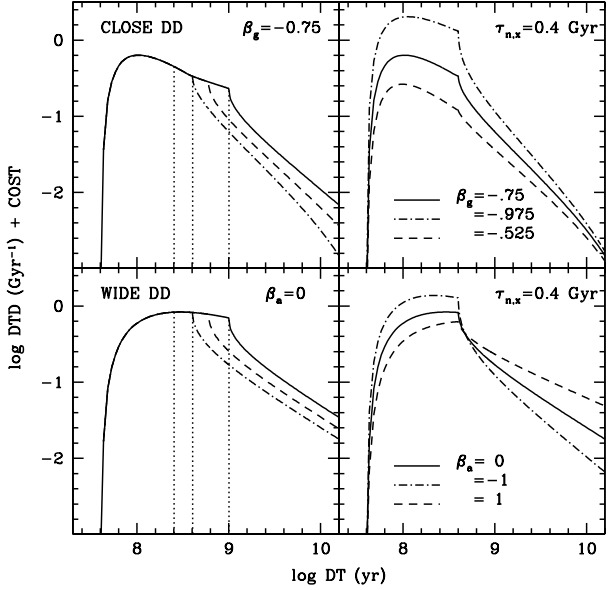}}
\caption{Effect of the main parameters on the DTD of DD CLOSE (upper panels) 
and DD WIDE (lower panels) models. The left panels show the dependence on 
the maximum evolutionary 
lifetime of the secondary: \taunx = 1 (solid), 0.6 (dashed) and 
0.4 (dash-dotted) 
Gyr, for the labelled values of \betaa\ and \betag.  
The right panels show the dependence on the distribution of the 
separations as labelled. Notice that the DTDs are not normalised to unity for
better illustration.}
\label{dd_param}
\end{figure}

\subsection[]{Double Degenerates}
  
Due to the relatively restricted mass range of the binary components (i.e. 
from 2 to 8 \msun) in the DD model, the DTD is not very sensitive to the 
IMF slope and to the distribution of the mass ratios, whereas the main 
parameters
controlling its shape are the maximum delay due to the MS evolution of the 
secondary (\taunx), and the shape of the distribution of the separations. 
The parameter \taunx\ is the evolutionary lifetime of the
least massive secondary in successful SN Ia progenitors: the heavier this
limit is, the earlier the position of the cusp in the DTD for both CLOSE and
WIDE DDs. The shape of the distribution of the separations of the DD systems
controls instead the overall slope of the decline as the delay time grows:
if most systems are born close, most events have a short \taugw\ and the 
DTD declines faster. Fig. \ref{dd_param} shows these dependences; notice that
the DTDs are not normalised to 1 for illustrative purposes.
The three values of \taunx\ which characterise the curves in the left panels
correspond to a minimum mass of the secondary in successful systems of 
2 (solid), 2.5 (dashed) and 3 \msun\ (dot-dashed): for normalised DTDs, 
the fraction of early
events is higher when this minimum mass is larger. 
The curves in the right panels all adopt a minimum \msec\ of 3 \msun, but
different distributions of the separations, as described by the parameters 
\betaa\ and \betag\ for the WIDE and CLOSE cases respectively. Although the two
parameters are defined in a different way (see Sect. 2.2), 
the smaller their values
the larger the fraction of systems with small separation \aff, and
the three curves  
correspond to a flat distribution (solid), a distribution more populated at 
small (dot-dashed) and at large (dashed) separations. For all the values of
the $\beta$ parameters considered here, the DTD declines towards long delay
times, but the slope is milder when \betaa\ or \betag\ are larger, 
since in this
case more systems have a long gravitational wave radiation delay. 
Consequently, the higher \betaa\ or \betag, the larger the fraction of 
events at late epochs.

\begin{figure}
\resizebox{\hsize}{!}{\includegraphics{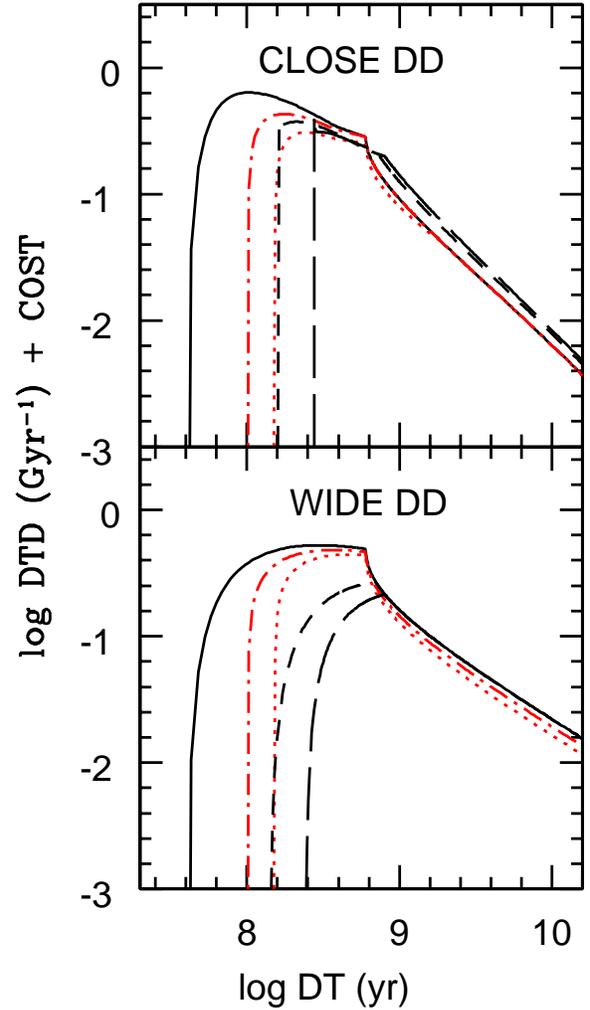}}
\caption{Effect of varying the minimum evolutionary lifetime (\tauni) and 
gravitational wave radiation delay (\taugi) in SN Ia progenitor systems for
DD CLOSE (top) and WIDE (bottom) models. The line-types encodes the values of 
(\tauni,\taugi) = (0.04,0.001) (solid), (0.1,0.001) (dot-dashed, red), 
(0.15,0.001) (dotted,red), (0.04,0.1) (short dashed), (0.04,0.2) (long dashed). All curves 
assume a minimum \msec\ of 2.5 \msun, and a flat distribution of \aff.} 
\label{dd_parama}
\end{figure}

Some other parameters have been fixed to derive the DTDs shown in 
Fig.~\ref{dd_param}: the minimum evolutionary lifetime of the secondary (\tauni)
 and the minimum \taugw\ (\taugi) in successful systems. The adopted
values are (\tauni,\taugi) = (0.04,0.001) Gyr, and correspond, respectively, 
to assuming that (i) the most massive secondary in SN Ia progenitors is an 
8 \msun\ star, and (ii) that the minimum delay due to gravitational wave 
radiation is negligibly small \footnote{The value of 1 Myr is imposed for 
numerical reasons.}. On the other hand, the upper limit to
\msec\ could be lower than 8 \msun, if, e.g., systems with massive 
secondaries feed the SD, rather than the
DD channel. Likewise, a significant lower limit to \taugw\ could be realised
in nature, if, e.g., DD systems which are born too close, or are too massive,
undergo an accretion induced collapse rather than a SN Ia explosion. Both 
alternatives deprive the
DTD of early events, while maintaining the population at long delay times.
Fig. \ref{dd_parama} shows the effect of varying separately 
\tauni\ and \taugi\ for the CLOSE and the WIDE DD models: some DTDs
assume a nominal \taugi\ coupled to \tauni=0.1 and 0.15 Gyr (corresponding to
a minimum \msec\ of 5.2 and 4.3 \msun\ respectively); some DTDs assume
the default \tauni=0.04 Gyr coupled to \taugi=0.1 and 0.2 Gyr. 
The latter limits
correspond to systems with a separation of 1 \rsun\ and masses \mdd=2 and 1.4
\msun\ respectively.  In all cases the
DTD is deprived of events with delay times shorter than (\tauni + \taugi), but
the overall shape is not much affected, though for the WIDE DD case 
a higher \taugi\ spreads the DTD toward longer delay times. It is also worth
noting that the cusp in all DTDs is actually located at \tage=\taunx + \taugi:
indeed, when the minimum gravitational wave radiation delay is non
negligible, the limit $\taun \leq \taunx$ corresponds to a discontinuity in the
parameter space for \tage\ located at \taunx+\taugi. 

\begin{figure}
\resizebox{\hsize}{!}{\includegraphics{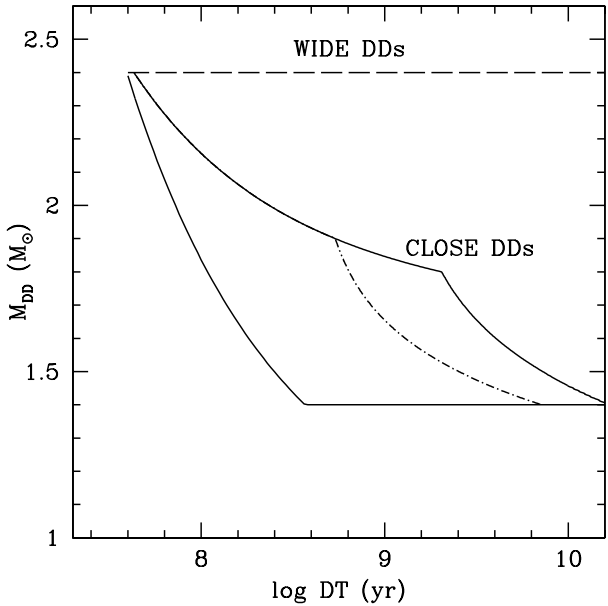}}
\caption{Illustration of the dependence on delay time of the range of DD 
systems total masses 
for the WIDE (dashed) and the CLOSE channels. For the latter, the upper limit 
shown as a solid 
(dash-dotted) line refers to a minimum secondary mass of 2 (3) \msun\ in SN Ia 
progenitors. A maximum \msec\ of 8 \msun\ and a negligible lower limit for
\taugw\ have been adopted. For delay times longer than $\sim 0.35$ Gyr 
the lower limit is given by the Chandrassekhar mass.}
\label{dd_syst}
\end{figure}

For the DD channel, at each delay time the events come from systems 
with a range of \taun\ (e.g. $\tauni \leq \taun \leq \tage - \taugi$)
and a corresponding range of \taugw. 
Therefore, systems with different masses of the
components and different separations end up exploding at the same delay time,
so that the occurrence of a significant discontinuity in the DTD 
at a particular delay time is unlikely.  
Similarly, there is no one to one correspondence between the delay time and
the mass of any of the WD components; however there is a trend of high mass
binaries providing preferentially explosions at early delays.

The diversity of the SN Ia light curves could be due to the range of
total mass \mdd\ if, e.g., the amount of \nira\ produced in the event was 
correlated to the total mass of the exploding system. For example, 
the extra mass $(\mdd\ - 1.4)$ could act as a {\it tamper} in the 
explosion, i.e.
inertially containing it a little longer, and lead to a higher production 
of \nira. Although this possibility has not been explored with theoretical
models, it is worth noting that the remarkable event SNLS-03D3bb was both 
over-luminous and 
Super-Chandrassekhar \citep{howell06}. Fig. \ref{dd_syst} 
illustrates the expected trend of the range of \mdd\
with the delay time. The lower limit represents the less massive
systems which have had time to evolve at a given \tage, i.e. binaries 
in which \msec\ is the evolutionary mass with lifetime equal to \tage,
and \mpri=\msec. For a negligible value of \taugi, some of these systems merge
at a delay time \tage. If the close binary evolution ends up with a wide range
of separations of the DD systems irrespective of the components' masses,
the maximum \mdd\ is the remnant of the (8+8) \msun\ systems,
i.e. \mdd = (1.2+1.2) \msun, independent of \tage.  
It will be in fact always possible to find a massive
system wide enough, so that the gravitational wave radiation delay
equals \tage-\taun. This upper limit could be appropriate for the WIDE DD case
(dashed line). However, massive binaries, which end up with massive remnants,
also undergo a severe shrinkage, especially in the CLOSE DD case. Therefore,
high \mdd\ will be associated to short \taugw, and short total delays in 
general; this introduces a trend for the upper limit to \mdd\ with \tage.
The solid line in Fig. \ref{dd_syst} illustrates this effect, and has
been computed by applying the \ace\ recipe to two consecutive RLOf episodes.
The exact location of this upper limit depends on various parameters (e.g. 
\ace, maximum separation of the interacting systems, initial-final mass
relation), but the general trend is realistic. In fact,
at each delay time, the maximum \mdd\ is realised in systems with maximum
remnants from \mpri\ and \msec\ which merge 
within $\taugw = \tage-\taun(\msec)$. 
Since \taugw\ is shorter in more massive systems, due to its explicit 
dependence on \mdd, as well as to the higher shrinkage, the above condition
is met for certain values of \mpri\ and \msec\ below their maximum
values of 8 \msun. The calculation shows that, as the total delay time grows, 
the condition above is first met with 
the maximum \mpri\ (8 \msun) in combination with a progressively less massive 
secondary, until the latter reaches the lower limit on \msec\ (e.g. 2 \msun)
adopted for the DD progenitors. 
At delay times longer than this the maximum \mdd\ is realised in systems with
\msec = 2 \msun\ and a progressively lighter maximum primary.
The transition between the two regimes 
is marked by the cusp in Fig.\ref{dd_syst}.

%Setting the primary mass to its maximum value, this condition is met with
%a certain value of \msec\ smaller than 8 \msun, since \taugw\ is shorter 
%in more massive systems, due to its explicit 
%dependence on \mdd, as well as to the higher shrinkage.
%As the delay times increases, the condition $\taugw=\tage-\taun(\msec)$ is
%met in systems with \mpri=8 and a progressively lower \msec. When the lower
%limit to \msec\ is reached, longer delay times require a relatively low
%mass primary. 
%At each delay time the maximum \mdd\ is realized in systems with maximum
%\mpri\ and \msec\ which merge within $\taugw \leq \tage-\taun(\msec)$;
%the delay \taugw\ decreases as \mpri\ and \msec\ increase, due to its
%dependence on \mdd, and to the higher shrinkage realized in more massive 
%systems. As the total delay time grows, the condition above is first met with 
%the maximum \mpri\ (8 \msun) in combination with progressively less massive 
%secondaries, until the latter reaches the lower limit on \msec\ (e.g. 2 \msun)
%adopted for the DD progenitors. Delay times longer than this are populated 
%with
%systems with the lowest possible secondary mass in combination with a 
%progressively less massive primary. The transition between these two regimes 
%is marked by the cusp in Fig.\ref{dd_syst}.

To summarise, for the WIDE DD case we expect a wide range of 
\mdd\ at any delay time, and virtually no trend for the average mass of the 
exploding WD with the
age of the parent galaxy. Instead in the CLOSE DD case we expect a definite
trend of decreasing average \mdd\ with increasing delay time, i.e. with
increasing average age of the parent population. In addition, the range
of \mdd\ in SN Ia progenitors should become narrower in old
systems, where the events should basically come from \mdd\ close to the 
Chandrassekhar mass.  

%At each delay time the maximum \mdd\ is realized in systems with maximum
%\mpri\ and \msec\ which merge within $\taugw \leq \tage-\taun(\msec)$: at
%short delay times this condition is met for \mpri=8 \msun in combination with
%an upper limit to \msec less massive than 8 \msun, due to the higher degree
%of shrinkage in systems with heavier secondary. By construction, this 
%condition can be fulfilled until \msec\ drops below its lower limit (e.g.
%2 \msun): when \tage\ grows longer than that the upper limit to \mdd\ is
%controlled by the upper limit on \mpri\ which ensures that the shrinkage
%is small enough to ensure long \taugw.

\begin{figure*}
\resizebox{\hsize}{!}{\includegraphics{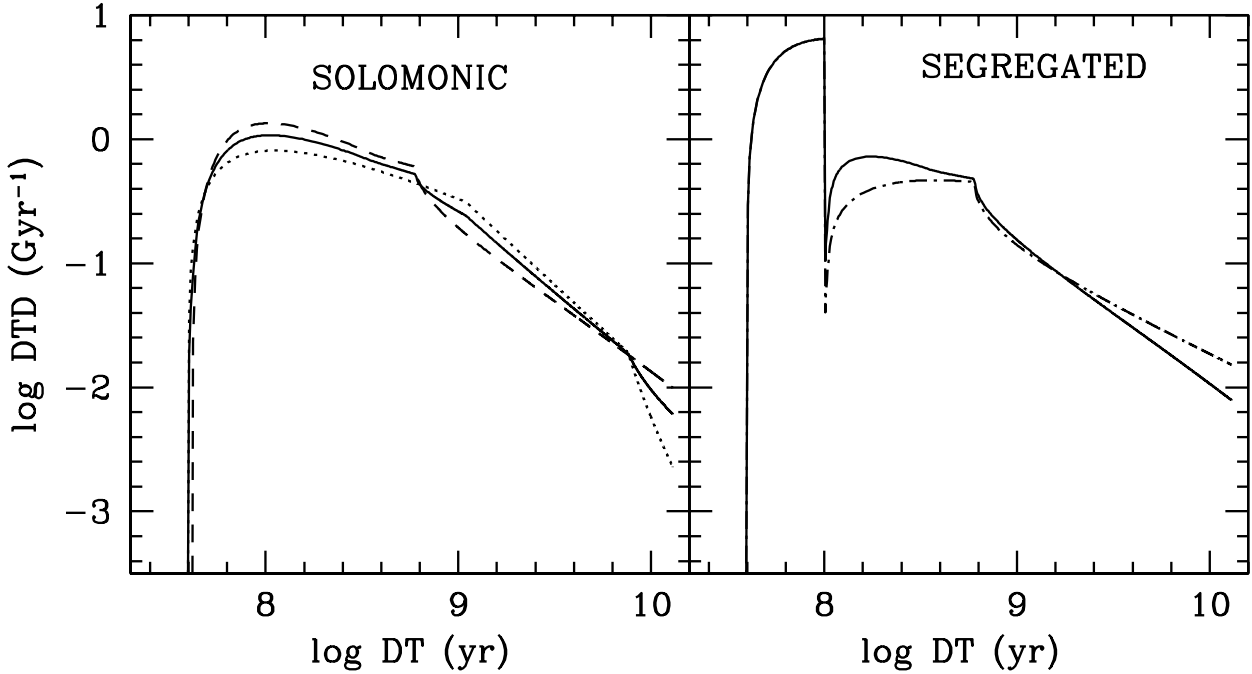}}
\caption{DTDs for mixed models. Left: the DTD of SD Chandra explosions 
in Fig. \ref{dtd_sd} (dotted line) is combined with the DTD of a DD-CLOSE
model with minimum secondary mass of 2.5 \msun\ and a flat distribution of
\aff\ (dashed) to construct the Solomonic mixture, 
in which 50 \% of the total
events within 13 Gyr come from either channel. Right: 
the events at delay times shorter than 0.1 Gyr come from SD explosions, 
while those at longer delay times come from the DD-CLOSE (solid) or 
DD-WIDE (dot-dashed) evolutionary channels, both with minimum \msec\
of 2.5 \msun\ and a flat distribution of \aff. The events from the SD channel 
amount to 30 \% of the total within 13 Gyr.}
\label{dtd_mix}
\end{figure*}

\section[]{Mixed Models}

As shown in the previous sections, the different channels for SN Ia production 
are in principle capable to provide both \textit{prompt} and \textit{tardy} 
events, as required by the data. Moreover, for all channels the DTD favours
early events, so that the SN Ia rate per unit mass in young stellar systems is
expected higher than in old ones, again as required by the 
observations. Although the fraction of early explosions is different for the
different channels, and within the same channel it depends on the 
astrophysical parameters which characterise the progenitor systems, each
model is compatible with the decreasing rate per unit mass going from
late to early type galaxies.
On the other hand, there's no real need to view the various models as mutually
exclusive, and both SD and DD paths could well be realised in nature.
From the stellar evolution point of view, the SD channel has more 
difficulty to produce Chandrassekhar explosions at very late epochs past star 
formation, because of the small mass reservoir from the donor; the DD channel
instead, can provide events at delay times exceeding the Hubble time due to
the possibility of realising very long \taugw. 
Therefore, rather than considering the two channels as mutually 
exclusive, it seems wiser to construct a mixed model, in which both SD and DD
events concur to the SN Ia explosion. Among the many possible combinations,
I consider two extreme hypothesis: the \textit{Solomonic} mix, in which both 
SD and DD channels contribute half of the total events within 13 Gyr; the
\textit{Segregated} mix, in which the SD explosions provide the early events,
while the DD channel feeds all the remaining delay times.
Although many other combinations are possible, it is instructive to see how 
these choices, which maximise the contribution of the two channels, impact on 
the DTD, and explore how the mix may be constrained by the observations.

Fig. \ref{dtd_mix} shows the Solomonic (left) and the Segregated (right) 
mixtures obtained with the particular choice of parameters for the
individual channels as described in the caption.
The Solomonic mix is characterised by a
DTD which is very similar to each of the SD and DD models used to construct it;
the only significant feature is the intermediate slope of the decline at
very late delays, where the number of events is kept relatively high due to
the contribution from the DD channel. This characteristic is relevant for the
SN Ia rate in the oldest galaxies, as will be illustrated later on. 
At each delay time, the ratio between DD and SD explosions varies,
but it keeps within a factor of 2 over almost the whole range of \tage:
only for $\tage \gtrsim$ 9 Gyr does the DD channel definitely prevail. 
This kind of mixture and its evolution over cosmic time has been studied in
\citet{grd}. In general, for the
Solomonic mixture, the majority of SN Ia events in early type 
galaxies should come from DD systems with \mdd\ close to the Chandrassekhar 
mass. The events in late type galaxies, instead, should come from both 
channels and may include super-Chandra explosions from DD progenitors.

\begin{figure}
\resizebox{\hsize}{!}{\includegraphics{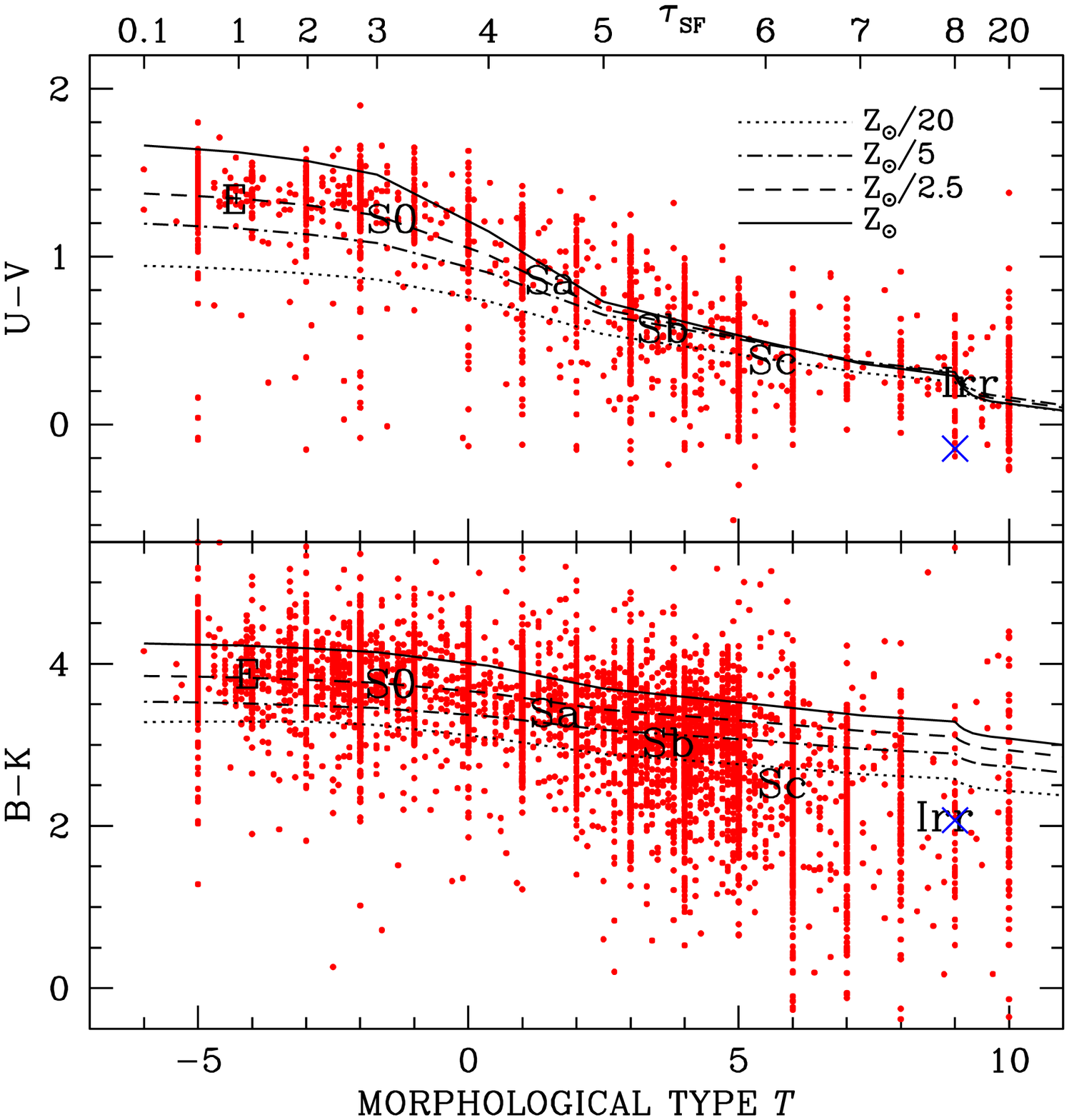}}
\caption{Calibration of the parameter \tausf\ of the SF rate versus 
morphological galaxy type. The red dots show the 
galaxies of the \citet{cet99} sample; the letters show the average location 
in color and type of the labelled galaxy class. The lines connect 
galaxy models 
obtained convolving the SF rate given by Eq. (\ref{eq_psi}) with the \tausf\
value labelled on the top axis (in Gyr) with simple 
stellar population (SSP) models. The latter were derived using the CMD tool 
(version 2.2) posted on the 
web (http://stev.oapd.inaf.it/cgi-bin/cmd), with 
the following input: \citet{marigo08} set of tracks; \citet{maiz} and
\citet{bessell} photometric system;
no circumstellar dust and no extinction; Salpeter IMF. The line type encodes
the metallicity of the SSP models, as labelled. The blue crosses shows the 
colors 
of a model with an exponentially increasing SF rate in the last 13 Gyr, and an
e-folding time of 1 Gyr, assumed to represent a bursting galaxy.} 
\label{calib}
\end{figure}

The Segregated mix assumes that systems with $\msec \geq 5 \msun$ feed the 
SD channel, while systems hosting less massive secondaries undergo CE 
evolution at the second RLOf, to become DDs, 
hence the discontinuity at 0.1 Gyr. This combination mimics the DTD 
proposed by \citet{mannu06} to explain the high SN Ia rate in radio 
loud Ellipticals as due to a recent SF episode in combination with a high
fraction of {\it prompt} explosions. Actually, if this were the case, the 
excess of 7 Ia events in the radio loud Ellipticals would imply
that $\sim$ 14 CC SNe should have been detected in the same sample, 
while none was observed \citep{grd}. It is anyway instructive to consider this
DTD as an example of a distribution with a high fraction of events at very 
short delay times; here a fraction of 30 \% \textit{prompt} events 
was chosen following the indications from chemical
evolution models by \citet{chicchi06}.
%loud ellipticals (but see \citealt{grd} for a criticism of this argument); 
%the fraction of
%30 \% \textit{prompt} events was chosen following the suggestion in
%\citet{chicchi06}. 
In this option, all events in early type galaxies come from 
the DD channel, while in galaxies with ongoing star formation we get a 
mixture of SD and DD explosions. In most cases, the two channels will be
providing events with similar probability: in fact the SN Ia rate in a 
galaxy $\sim$ 13 Gyr after the beginning of star formation is given by:

\begin{equation}
\dot{n}_{Ia}(13) = \kaia \times \left( \psi_{\rm c} \int_{0}^{0.1} \fia(\tage)\,\rmn{d} \tage + \psi_{\rm p} \int_{0.1}^{13} \fia(\tage)\,\rmn{d} \tage \right) 
\label {eq_rate}
\end{equation}

\acap
where \kaia\ is the number of SN Ia from a stellar population of unitary mass
(or SN Ia productivity), and $\psi_{\rm c}$ 
and $\psi_{\rm p}$ are the (\fia\ weighted) average SF rates 
over, respectively,
the last 0.1 Gyr and from 0.1 up to 13 Gyr ago.
Equation (\ref{eq_rate}) explicitly shows the two contributions of 
\textit{prompt} and \textit{tardy} components of the DTD: if their ratio 
is of 0.3 to 0.7,
in order to have an equal probability to realise SD or DD explosions it
suffice that the current SF rate exceeds the average SF rate in the past by a 
factor of $\sim$ 2.

To summarise, both the Solomonic and the Segregated mixtures provide SD and DD
explosions in virtually all galaxy types, but in the very old galaxies, where
(almost) all events should come from the DD channel. It is worth noting that 
this conclusion only applies to Chandra explosions, since Sub-Chandra events 
could be common at very late delay times both in the Single and Double 
Degenerate scenarios.  

The scaling of the SN Ia rate in different stellar systems 
offers the opportunity of constraining the DTD, hence the SN Ia progenitor 
model (e.g. Paper I,\citealt{mannu06,blanc}). To this end I consider two sets 
of data: (i) the relation between the SN Ia rate per
unit luminosity and the morphological type of the parent galaxy 
(\citealt{gc09}, hereinafter GC09) 
and (ii) the relation between the SN Ia rate per unit mass and the specific SF
rate (SSFR) of the parent galaxy \citep{sulli06}. 
Both relations (i) and (ii) basically stem from DTD being more populated 
at early delay times: as
a consequence, the rate per unit mass is higher the younger the stellar 
system. A later galaxy type and/or a higher specific SF rate
correspond to a larger fraction of young stars, hence a higher SN Ia rate per 
unit mass, or per unit luminosity, as far as luminosity traces the stellar 
mass. Nonetheless the two observational relations are largely independent 
(i) because of the different galaxy sample, and (ii) because of the different 
normalisation of the SN Ia rate, and different galaxy age tracer. Therefore,
they yield complementary information on the DTD. 
%The quantitative slope of the correlations depend on the shape of 
%the DTD, and on the SF history in the various galaxies. 
In the next sections, the 
quantitative fit of the two relations is attempted separately.  
%This issue is addressed
%in the next sections separately for the two observed relations.
%
%Therefore, \citet{sulli06} data yield 
%an additional constraint on the shape of the DTD. 

%\subsection[]{Constraining Mixed Models}
%\subsection[]{Constraining Mixed Models: rates as function of galaxy type}

\subsection[]{Rates as function of galaxy type}

%The correlation between the SN Ia rate per unit luminosity and the 
%morphological type of the parent galaxy is 
%equivalent to the relation between the SN Ia rate per unit luminosity and
%the color of the parent galaxy, but it allows us to take advantage of the
%whole sample of events in \citet*{cet99}, as well as to  

The indication that late type galaxies are more efficient in producing SN Ia 
events goes back to \citet{tammann}, and was interpreted as due to a DTD more
populated at short delay times already by \citet{oemler} and \citet{gr83}. 
Since then, this observational feature has been put on firm grounds as
a correlation between the SN Ia rate per unit B-band luminosity (rate in
SNu's \footnote{Supernovae rate units are SNu's, SNuK's and SNuM's, 
corresponding to events per century per $10^{10}$ \lbsun, \lksun\ and 
\msun\ respectively.}) with the $U-V$ 
colour of the parent galaxy \citep*{cet99}, and a correlation between the
rate per unit K band luminosity (rate in SNuK's) and the $B-K$ colour of the 
parent galaxy \citep{mannu05}. Both correlations are based on the 
\citep{cet99} sample of (nearby) galaxies. \citet{mannu05} also derived 
the rate per unit mass (rate in SNuM's) as a function of the parent galaxy 
$B-K$ colour, later used in Paper I and in \citet{mannu06} to
derive clues on the characteristics of the DTD. In particular, Paper I found that
both CLOSE and WIDE DD models could explain the data, provided that the DTD
slope was neither very flat nor very steep. However, 
the rate in SNuM's relies on theoretical models which are
needed to convert the galaxies K-band luminosities into their mass; therefore
the purely {\it observed} relations involve the rates per unit luminosities.
The attempt to account for both \citet{cet99} and \citet{mannu05} 
trends of the rates in SNu's and SNuK's with the parent galaxy $U-V$ and 
$B-K$ colors fails (GC09), with the
rate per unit \lb\ requiring a flatter DTD than the rate per unit \lk.
This is because the slope of the observed correlations depend on both the
DTD and on the SF history assumed to model the galaxies. To improve on the
description of the latter, GC09 constrained the sample galaxies to
be represented by the family of SF history laws adopted by \citet{gavazzi}: 

\begin{equation}
\psi(t) = \frac {t}{\tausf^2} \, exp \left( \frac{-t^2}{2 \, \tausf^2} \right)
\label{eq_psi}
\end{equation}   

\acap
which peaks at progressively older ages as the parameter \tausf\ decreases.
A short \tausf\ also implies a narrow age distribution, akin to early type 
galaxies, while for $\tausf \ge 13$ Gyr the SF rate monotonically increases 
up to the present, having adopted an age of 13 Gyr for all galaxies.
A calibration of \tausf\ versus the morphological galaxy type $T$ (as given
in the RC3 catalogue, \citealt{devau})
was obtained by fitting the $U-V$ and $B-K$ colours of the sample galaxies.
In this way theoretical SN Ia rates (in SNu's and in SNuK's) as a
function of $T$ could be computed and compared to the measured correlations.
Such procedure has two advantages: the galaxies' SF history
is constrained from their luminosity in four photometric bands, and the full
sample of SN Ia events in \citet{cet99} can be used to derive the rates in 
SNu's because all galaxies have a determined \lb\ and type $T$, 
while \lk\ is available for only a fraction of them. 

Fig. \ref{calib} shows the calibration of the \tausf\ parameter 
versus $T$: the average colors of E, S0, 
Sa, Sb and Sc galaxies are reproduced with respectively \tausf $\simeq$ 1,3, 
4.5, 5.5 and 6 Gyr. It also appears that late type galaxies are better 
reproduced with a lower metallicity than early types. 
Concerning Irregular galaxies, the models account well for their average 
$U-V$ color, but appear too red in $B-K$ compared to the data. 
This may result from an inadequate 
description of their SF history, which could be dominated by a current burst.
The cross in Fig. \ref{calib} shows the location of a bursting model, which
well reproduces the $B-K$ color of this galaxy type, but turns out
too blue in $U-V$, hence not providing a satisfactory fit either.
For all other galaxy types, however, the average colors are well reproduced by
the adopted SF history law, which is used in the following to fit the 
correlation of the SNIa rate with $T$.

\begin{figure*}
\resizebox{\hsize}{!}{\includegraphics{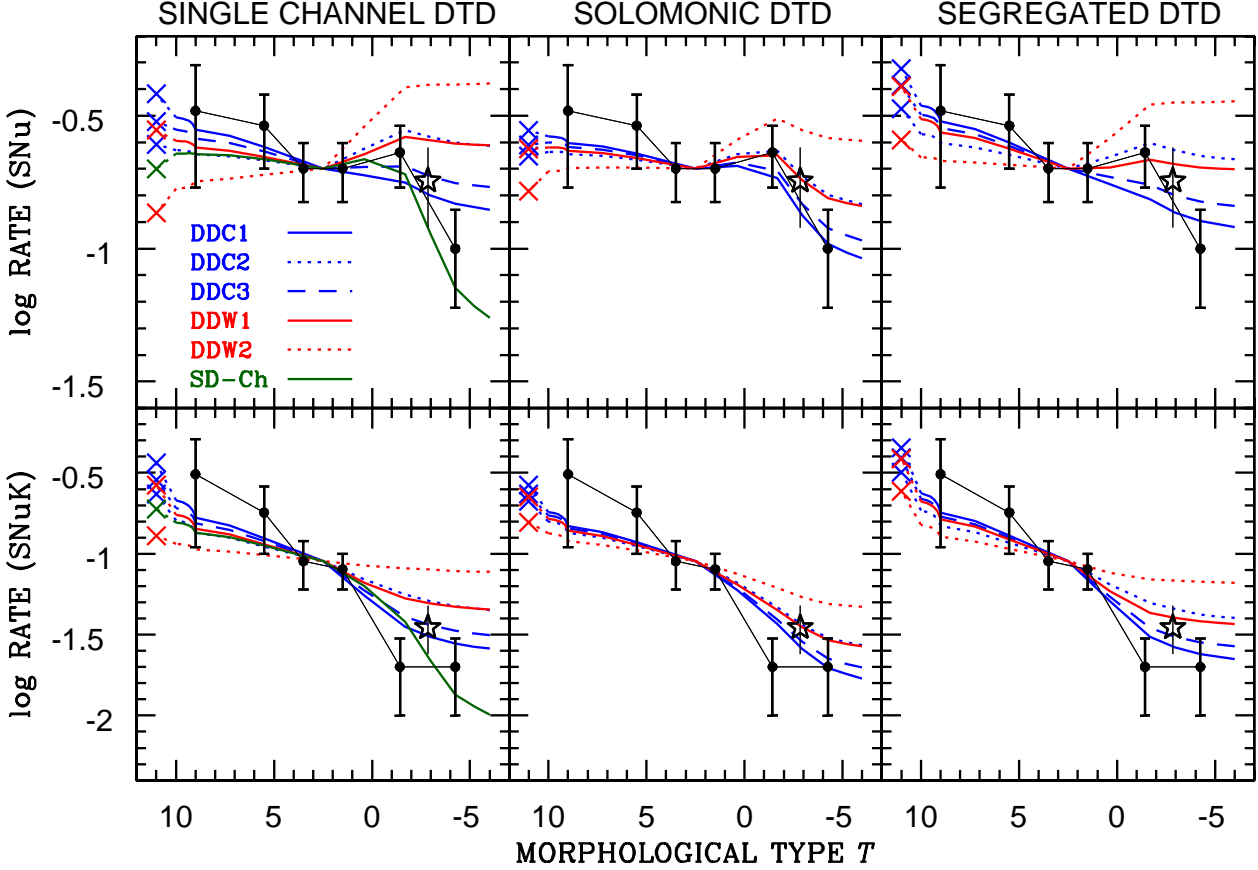}}
\caption{SN Ia rate per unit B (upper panels) and per
unit K (lower panels) luminosity, as function of the morphological 
type of the parent galaxy. 
Black dots show the observed rates from \citet{gc09};
the stars shows the published rates for the E+S0 class;
coloured lines show models calculated with different DTDs and the SF rate 
given by Eq.(\ref{eq_psi}). Left panels: theoretical relations obtained with 
single channel DTDs.  The SD Chandra (SD-Ch,m green line) model assumes a 
Salpeter distribution for \mpri, a flat distribution of the mass ratios, and 
an efficiency $\epsilon=1$. Blue lines are obtained with DD-CLOSE models; 
DDC1 and DDC3 assume a distribution of separations skewed at small values,
and differ for the minimum \msec, respectively of 2.5 and 2 \msun; DDC2
assumes a flat distribution of the separations and a minimum \msec\ of
2.5 \msun. Red lines are obtained with DD-WIDE models, DDW1 and DDW2 being
computed with the same choice of parameters as DDC1 and DDC2.
Central and right panels: correlations obtained with mixed DTDs in the 
Solomonic and Segregated flavours; colours and line types reflect the 
specific DD model used to construct the mixture with the encoding
labelled in the top left panel. The coloured crosses show 
the rates for the bursting model in Fig. \ref{calib} put at an arbitrary
value of $T$.}  
\label{morph}
\end{figure*}

The black dots in Fig. \ref{morph} shows the SNIa rates as function of
morphological type determined in GC09, while the star
shows the rate determined for the merged class of early type galaxies 
(E and S0) in \citet{cet99} (upper panel) and \citet{mannu05} (lower panel).
The latter value turns out higher than the rates determined by GC09 
in the individual classes; this is mostly due to a (slightly) different 
binning of the parameter $T$ which identifies the morphological types. 
%due to different criteria used for
%coincidence of 2MASS and RC3 galaxies, and to the (slightly) different 
%binning of the morphological type T. \citet{mannu05} perform
%a source-source coincidence with a tolerance radius of 20 \", while 
%\citetGC09 use NED to identify the source by its name. This results into
%a sample of 2048 E+S0 galaxies hosting 21 SN Ia in \citet{mannu05}, versus a
%sample of 2285 E+S0 galaxies hosting 10 SN Ia in \citetGC09. 
Coloured lines in Fig. \ref{morph} show theoretical relations obtained with 
selected DTD models (see caption), meant to show the 
dependence on the main parameters. The models are computed adopting the 
correspondence between \tausf\ and $T$ just discussed, plus a relation between
\tausf\ and metallicity $Z$, with $Z$ decreasing from
0.008 to 0.001 as $\tau_{SF}$ goes from 1 to 20. 
The crosses in Fig. \ref{morph} show the location of the bursting model (cross
in Fig. \ref{calib}). 
%models calculated with an exponentially increasing SF rate from 13 Gyr ago up 
%to now, with an e-folding time of 1 Gyr. 

It is worth emphasising that,
rather then simulating a large population of individual objects, the approach
here aims at describing the photometric properties and the SNIa rate of the
\textit{average} galaxy within each bin of $T$; the morphological type is
assumed to trace the age mixture in the galaxies, and describe its changes 
across the  Hubble sequence.

The theoretical rates in Fig. \ref{morph} are computed as the product of 
the rate per unit mass and the mass-to-light ratio of the galaxy models:

\begin{equation}
\frac{\dot \nia(t)}{L_{B(K)}} = \kaia \, \frac{\int_0^{13}\psi(13-\tau) \fia (\tau) \rmn{d} \tau}{\msf} \times \frac{\msf} {\int_0^{13}\psi(13-\tau) \lbk (\tau) \rmn{d} \tau}
\label{eq_nialum}
\end{equation} 

%\begin{equation}
%\frac{\dot \nia(t)}{L_{B(K)}} = \kaia \, \frac{\int_0^{t}\psi(t-\tau) \fia (\tau) \rmn{d} \tau}{\int_0^{t}\psi(t) \rmn{d}t} \times \frac{\int_0^{t}\psi(t) \rmn{d}t} {\int_0^{t}\psi(t-\tau) \lbk (\tau) \rmn{d} \tau}
%\label{eq_nialum}
%\end{equation} 

%\begin{equation}
%\frac{\dot \nia(t)}{L_{B,K}} = \kaia \, \frac{\int_0^{t}\psi(t-\tau) \fia (\tau) \rmn{d} \tau}{\int_0^{t}\psi(t-\tau) \lbk (\tau) \rmn{d} \tau}
%\label{eq_nialum}
%\end{equation} 

\acap
where \lbk\ are the light-to-mass ratios of simple stellar populations, 
either in the B or in the K
band, as functions of age, and \msf\ is the total mass transformed into stars
(i.e. integral of $\psi(t)$) up to the current galaxy age, here assumed of 
13 Gyr. 
When going from early to late type galaxies, the increase of the first factor
(rate per unit mass) competes with the decrease of the second factor
(mass to light ratio), which is more pronounced in the B, rather than K, band.
This explains the large variation of the model rate in SNuK's compared to 
that in SNu's; it also explains why the rate in SNu's can even be 
predicted higher in early than in late type galaxies (see e.g. model DDW2). 

%Although the rate per unit mass increases going from early to late type
%galaxies, younger populations have lower mass-to-light ratios, which 
%counteracts
%the trend of the rate per unit mass. This effect is stronger in the $B$ 
%than in the $K$ band. 
The models are normalised to the observed value at 
$T =3$, intermediate between Sa and Sb galaxies; the normalisation yields a
value for \kaia, i.e. the productivity of SN Ia from one unit mass stellar
population, which depends on the assumed IMF. The observed rates imply
$\kaia \simeq 10^{-3} \msun^{-1}$ with relatively little dependence on the 
DTD or  the photometric band, for the Salpeter IMF adopted in the 
CMD tool (i.e.
$\phi(m) \propto m^{-2.35}$ down to 0.01 \msun). A bottom light IMF, with
$\phi(m) \propto m^{-1.3}$ between 0.1 and 0.5 \msun\ and a Salpeter slope
above 0.5 \msun, would yield \kaia\ values larger by a factor of $\sim 3$.
A detailed discussion of the uncertainty on \kaia\ requires a thorough 
evaluation of the quality of the fits, which goes beyond the scope of this 
paper. It is however interesting to notice that this value of the
SNIa productivity, compares well with the total Fe content of galaxy clusters. 
According to \citet{alvio04} the Fe Mass-to-Light ratio of clusters is 
$M_{\rm Fe}/L_{\rm B} \sim 0.015$ in solar units, assuming a 
stellar mass-to-light ratio of $M_*/L_{\rm B} \sim 3.5$. This value
is appropriate for a bottom light IMF, for which the current mass in stars
(at old ages) is $\sim$ 0.6 of the total mass transformed into stars.
Therefore, the Fe content in galaxy clusters requires an Fe productivity of 
$M_{\rm Fe}/\msf = 0.015 / (3.5/0.6) = 0.0026$. Adopting 
$\kaia = 0.003 \msun^{-1}$ and an Fe production of 0.6 \msun\ per SNIa
event, the Fe productivity from SNIa's is of 0.0018, which accounts for $\sim$
70 \% of the Fe content in galaxy clusters; the remaining 30 \% would be
produced by CC SNe.

Fig. \ref{morph} shows that the models with a flat distribution of the
separations (dotted lines) tend to over predict 
the rate in Ellipticals, in both DD CLOSE and DD WIDE cases.
This problem is lessened for the Solomonic mixture, due to the
dramatic drop of the SD channel at long delay times. 
However, when considering the merged E+S0 class, the
DD CLOSE model with a flat distribution of \aff\ becomes compatible with
the data. It is then very important to derive a robust measurement of the
SN Ia rate separately for the two morphological types:
in early type galaxies, a difference in average age of $\sim$2 Gyr has 
a strong effect on the SN Ia rate per unit mass, if a sizable contribution
from the SD channel exists. The fact that the rate in
SNu's in S0 galaxies appears so much higher than in Es favours this 
interpretation, although the statistics is still too poor to draw such
conclusion. In this respect, the SD model appears to best represent 
the data. However, the green lines in Fig.~\ref{morph} are obtained with 
a choice for the accretion efficiency and distribution of the mass ratios 
which maximize the fraction
of systems with long delay times; any other choice would lead to an 
underproduction of SNIa events in E galaxies.

Considering now the rates in late type galaxies, all models are consistent
with the data within the very large error bar, although the predicted rates
are lower than the average observed values. 
The parameter \taunx\ which has some impact on the fraction of events at
short delay times (see Sect. 3.2) does not appear to play a role in the fit
(e.g. DDC1 vs DDC3). 
In a star bursting regime the SN Ia rate gets substantially 
increased (crosses), especially when the DTD is particularly skewed towards the
short delay times. 
As mentioned above, the galaxy modelling 
presented here does not reproduce satisfactorily the colors of Irregulars with
either tested shapes for the SF rate.
A more precise assessment of the recent SF in 
galaxies of the latest types is crucial to derive information on the DTD 
from their observed SN Ia rates.

As a general conclusion, the trend of the SN Ia rate as function of the 
morphological type suggests that the DTD ought to be significantly 
decreasing  at the very long delay times. This disfavours flat distributions
of the separations in the case of DD progenitors. Provided that, the fit 
obtained with the mixed models is not particularly better than that with
single channel models.  

\begin{figure*}
\resizebox{\hsize}{!}{\includegraphics{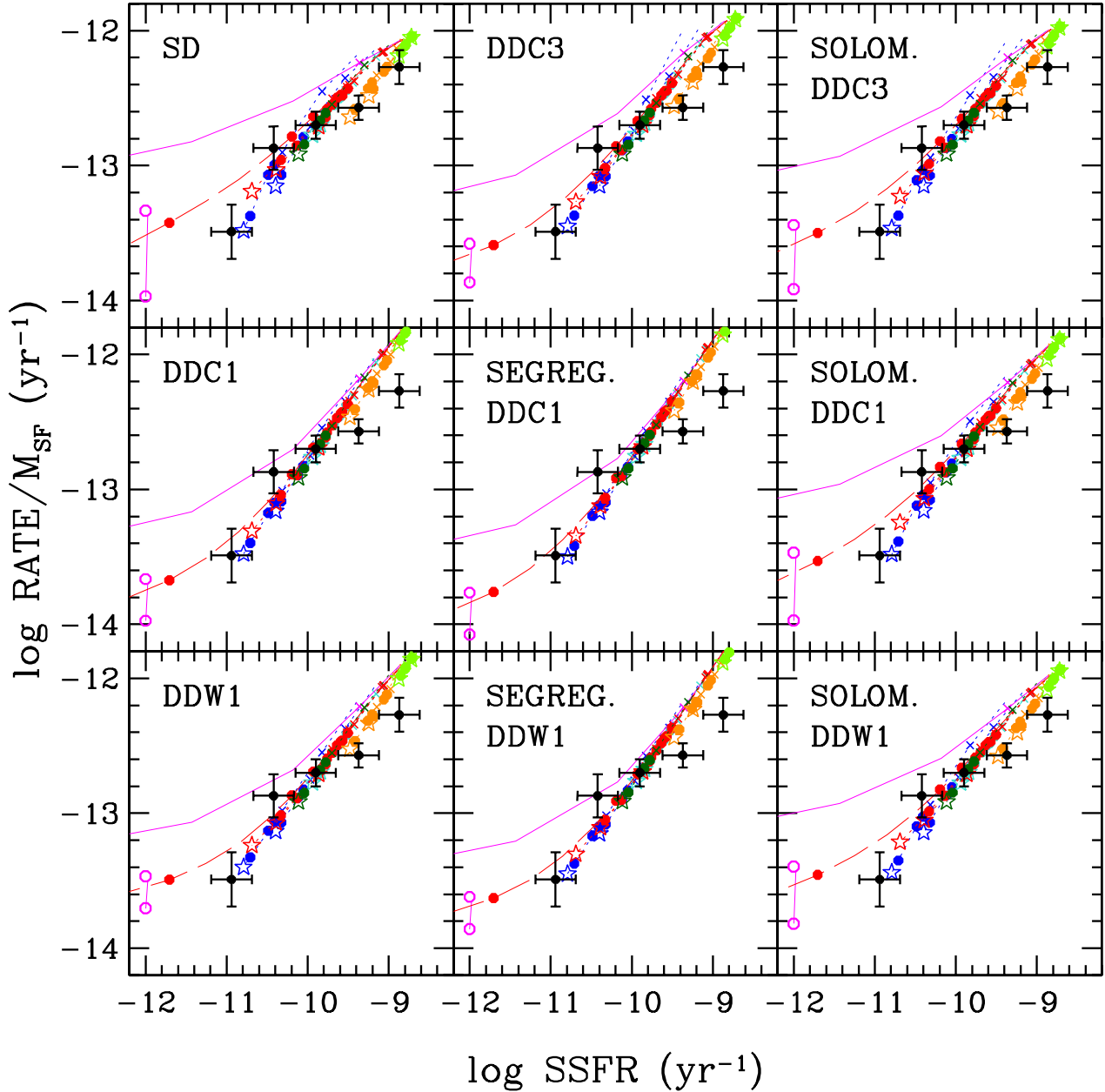}}
\caption{Correlation between the SN Ia rate and the SF rate both normalised to 
the total mass transformed into star. 
The data (black dots) are from \citet{sulli06} shifted down and left to 
account for the mass return, by (0.12,0.12,0.15,0.17,0.19) dex for the 5 bins 
from right to left. The models (coloured lines and dots) assume 
different DTDs as labelled in each panel (see caption to Fig.\ref{morph} for
the meaning of the acronyms). The colours encode the kind of SF history:
(i) currently bursting, with a ratio of current to past SF rate of 500, 50 
(light green), 10, 5 (orange); (ii) power laws with exponent -0.9 and -0.5 
(blue), 0.5 and 0.9 (cyan); (iii) $\psi$ given by Eq. (\ref{eq_psi}) 
with $\tau_{SF}$=1 (magenta), 3, 5, 6 and 20 Gyr (red).
The lines show the models as they age from 1 to 13 Gyr; crosses show the 
location at 2 and 5 Gyr, full dots at 6, 7 and 11 Gyr, the stars mark the
position at 13 Gyr. The two open dots show the rates of the model with $\psi$
given by Eq. (\ref{eq_psi}) and $\tau_{SF}$=1 Gyr at an age of 7 (upper) and
11 (lower) Gyr. Notice that the SSFR of these models is much lower than 
$10^{-12}$ yr$^{-1}$, and virtually null.} 
\label{pritchet}
\end{figure*}

\subsection[]{Rates as function of specific star formation}

\citet{sulli06} measured the SN Ia rate per unit mass as a function of the 
current star formation rate per unit mass for the Supernova Legacy Survey
(SNLS) galaxy sample \citep{sulli05}. The correlation is based on 124 SN Ia 
at redshifts between 0.2 and 0.75.
It is worth recalling that in this work the galaxy masses were determined 
via spectral energy distribution (SED) fitting
to P\'EGASE 2.2 models \citep{rocca97} with \citet{kroupa} IMF, and that the 
SSFR is measured as the ratio between the mean SF rate over the last 
0.5 Gyr and the current stellar mass of the galaxy, resulting from the 
SED fitting.
%As already pointed out, the correlation of increasing SN Ia rate per unit 
%mass with increasing SSFR results from the DTD being more populated at short 
%delay times, i.e. the same property at the basis of the correlations described
%in the previous section. 
%However, the two observational relations are 
%substantially independent (i) because of the different galaxy sample; 
%(ii) because 
%of the different normalization of the SN Ia rate and different galaxy age 
%tracer. Therefore, \citet{sulli06} data yield an additional constraint on the
%shape of the DTD. 

Black dots in Fig. \ref{pritchet} show the observed relation for the star
forming galaxies in \citet{sulli06}; coloured dots and lines are models 
constructed for a wide variety of SF histories, so as to illustrate
the impact on the theoretical relation of the adopted SF law, and of the 
galaxy age. Given their redshift range, the SN Ia host galaxies should have
current ages between $\sim$ 7 and 11 Gyr, or younger. 
Technically, the theoretical specific SN Ia and SF rates are computed as:

\begin{equation}
\frac{\dot \nia(t)}{\msf} = \kaia \, \int_0^{t}\psi(t-\tau) \fia (\tau) \rmn{d} \tau
\label{eq_niamass}
\end{equation} 
 
\begin{equation}
\frac{\dot M}{\msf} = 2 \, 10^{-9} \, \frac{\int_{t-0.5}^{t}\psi(t) \rmn{d} t}
{\int_{0}^{t}\psi(t) \rmn{d} t}
\label{eq_ssf}
\end{equation} 

\acap
where $t$ is the current age and \msf\ is the mass transformed into stars 
up to now. Because of the mass return from dying 
stars, \msf\ is always higher then the 
current stellar mass in a galaxy, by a factor of $\sim$ 1.6, 1.3 for old and 
young stellar populations, respectively. These values hold for a bottom 
light IMF, such as Kroupa's, and for galaxy ages older than $\sim$ 1 Gyr. 
Therefore, accounting for the mass return corresponds to shifting the models 
up and right by $\sim 0.1 - 0.2$ dex, the correction being however
different for each model, since it depends on age and SF history.
Given the differences in the galaxy
modelling between \citet{sulli06} and this paper, rather than correcting each 
model, the data points (read off Fig. 3 in \citealt*{pri}) have been shifted
by a variable amount (see caption). The applied correction is meant 
to represent the
average ratio of the current stellar mass to the total mass gone into stars
in the five bins. 
As usual, the productivity \kaia\ in Eq. (\ref{eq_niamass}) is a free 
parameter; however the models in Fig.\ref{pritchet} have been all computed
with \kaia = $10^{-2.8} \msun^{-1}$ irrespective of the DTD, a value which 
turns out to describe very well the SN Ia rate in the bin at 
SSFR=$10^{-10}$ yrs$^{-1}$. A refined fit of the SN Ia productivity seems 
unworthy, since the models and the data are not completely compatible as 
just discussed, and since the SF history of the galaxies is not well 
constrained. 

The slope of the theoretical correlation between the rate per unit mass and 
the SSFR appears virtually insensitive to the assumed SF history, 
for star forming galaxies, as already pointed out by \citet{pri}. 
The models overlap each other along a locus which
is well described by a power law along which 
the younger the galaxy (either because of its current or its average age), 
the higher its SSFR, \textit{and} its SN Ia rate per unit mass.
The exponent of the power law is sensitive to the model DTD; for the models
plotted in Fig.\ref{pritchet} it ranges from 0.6 (SD model) to 
0.8 (Segregated mixture with DDC1 model). In addition, the theoretical 
relation is found systematically 
steeper for the Segregated mixtures compared to Solomonic mixtures or 
Single channel models.
 
In general, all DTDs yield and acceptable representation of the data.
Taken at face value, the observed correlation seems to favour models with
a moderate power at the early delay times, as the SD model, or the Solomonic
mixtures. The Segregated mixtures tend to predict a
too high SN Ia rate in star bursting galaxies at given SSFR value, but
no strong conclusion can be drawn from this comparison.
\citet{sulli06} also determine the SN Ia rate per unit mass in
\textit{passive} galaxies (from SNLS data), which turns out very close to 
the rate measured for galaxies in the bin at SSFR=$10^{-11}$ \msun/yr.  
This point is not plotted on Fig.\ref{pritchet}, where, instead, I show the 
location of two models for old stellar populations (see caption), in which
most of the stellar mass has been made at high redshift.  
%obtained with a SF history described by 
%Eq.(\ref{eq_psi}) adopting $\tau_{\rm SF}$=1 Gyr,
%at ages of 7 and 11 Gyr. 
%These models have virtually null current specific SF rate,
%because most of their stellar mass has been made at high redshift. 
These models have virtually null current specific SF rate and yet 
their current SN Ia rate is relatively high due to the late delay times 
component of the DTD. 
This interpretation differs from what concluded in \citet{pri} mainly
because of the different rendition of the SF history in passive galaxies.
Adopting Eq.~\ref{eq_psi} with a short \tausf, the SSFR quickly drops 
as the time from the beginning of star formation increases, so that negligible
SSFR are obtained when the SNIa rate is still relatively high because of the
\textit{tardy} component. For example, a model with $\tausf =2$ Gyr, at an
age of 10 Gyr has a SSFR of $1.8\, 10^{-14}$ \msun/yr and an average age
of 7.5 Gyr. With this star formation history, the SD DTD plotted in 
Fig. \ref{pritchet}
predicts a SNIa rate of $4.7\, 10^{-14}$ (\msun yr)$^{-1} $, matching the 
rate in passive galaxies determined by \citet{sulli06}. 
%When allowing for a prolonged SF episode, albeit at old ages, even the SD 
%model can maintain a relatively high SN Ia rate. 
However, as noticed in the previous section, such SD model assumes a 
100 \% accretion of the envelope of the secondary onto the CO WD, and a flat 
distribution of $q$, so that the rate at late epochs is enhanced. 
For this reason, DD models, and/or Solomonic mixtures could be more plausible.

%\begin{figure}
%\resizebox{\hsize}{!}{\includegraphics{prompt.eps}}
%\caption{Fraction of events with delay time less than 0.3 Gyr as a function of
%the ratio between the average SF rate over the last 0.3 Gyr and that over the
%previous 12.7 Gyr, for the DTDs used in Fig. \ref{morph} with the
%same line encoding as in Fig. \ref{cumul}}.
%\label{prompt}
%\end{figure}

\begin{figure}
\resizebox{\hsize}{!}{\includegraphics{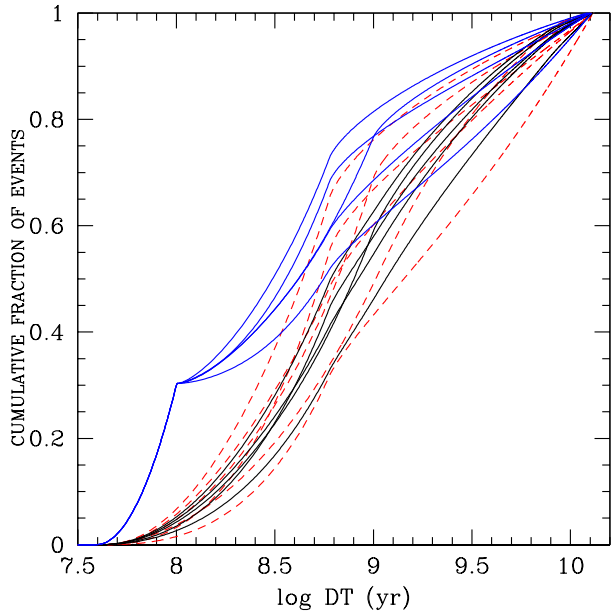}}
\caption{Cumulative fraction of events as the delay time grows for 
the DTDs used in Fig. \ref{morph}. Dashed red, solid black and
solid blue lines refer respectively to single channel, 
Solomonic mixtures and Segregated mixtures.}
\label{cumul}
\end{figure}

\section[]{Summary and Discussion}

%general philosophy: concept of the paper and how the questions are tackled
This paper presents the theoretical expectations for SN Ia progenitors 
%based on broad stellar evolution arguments, 
focusing on two questions: the shape of 
the distribution of the delay times and the possible diversity of SN Ia 
progenitors of events at different delay times. These two 
issues impact on the cosmological use of SN Ia's as distance indicators because 
they may entail a systematic diversity between events at low and high 
redshift, where the progenitors are younger and more massive.
As in Paper I, the approach refrains from describing the evolution of individual 
systems, rather concentrating on the clock of the explosion and on a few 
parameters singled out as crucial for the two issues. In particular, 
different from BPS renditions, the mass exchange phases during the binary 
evolution are not
followed, and their result in promoting or not the final event is considered 
as an equally plausible option. The dependence of the likelihood of the SN Ia 
event on the 
initial binary separation is also neglected, having assumed that (i) there is
a range of separations suitable for the realisation of the SN Ia, and that
(ii) this range does not depend on the other binary parameters 
(\mpri\ and \msec). While strictly speaking the latter assumption is untrue
(see discussion at the beginning of Section 2.3), the 
comparison of the analytic DTDs with the results of BPS simulations seem to
indicate that the crucial parameters are indeed the masses of the components.  

% basically only Chandra models are investigated.Sub-Chandra however could have
% great potential in old SPs (both for the SD and for the DD cases)
%main features of single channel models: parameters, comparison with BPS
% prompt and tardy, discontinuities
I have illustrated the main features of the analytic DTD for the single and 
the double degenerate evolutionary channels, concentrating on Chandrassekhar 
exploders,
since Sub-Chandra models are currently disfavoured. Nevertheless, it is
worth pointing out that Sub-Chandra events could be important especially to 
account for SN Ia's in old stellar populations. Actually these events are an
embarrassment, since they are much easier realised than Chandra explosions,
%many are predicted to occur, 
yet there are no established observational counterparts.
Crucial parameters for shaping the SD DTD are the limits on the primary 
and on the secondary mass, and the efficiency of accretion onto the CO WD. 
The DTD is populated only within delay times comprised between the MS lifetime 
of the most and least massive secondary in the progenitor's system; a lower
accretion efficiency results into a steepening of the DTD at late epochs.
If only half of the envelope of the secondary is used for the mass growth of
the CO WD, the DTD drops by three order of magnitudes between 1 and 10 Gyr,
making the rate in old stellar systems negligibly small.
For both flavours of DD models (CLOSE and WIDE), the major parameters are the 
minimum mass of the secondary in successful systems and the slope of the 
distribution of the separations of the DD systems at birth. 
Additional parameters are the minimum values for the delays \taun\ and \taugw,
the former being directly connected to the maximum value of \msec\ in
successful systems. 
The comparison of the analytic DTDs with results in the literature shows a very
good agreement with BPS renditions for the DD CLOSE channel (renditions for 
the DD WIDE channel are lacking in the literature). Less favourable is the
comparison for the SD channel, for which the BPS results yield a large variety
of DTD shapes. Clearly, in the simulations, the efficiency with which this 
channel produces SN Ia 
events is very sensitive to the criteria adopted to describe the mass 
exchange phases and the fate of the accreting WDs.

\subsection[]{Prompt and tardy components}

In principle, all channels accommodate {\it prompt} and {\it tardy} events, 
but the DTD is
very continuous so that this distinction appears artificial. As 
illustrated in Fig. \ref{cumul}, for the single channel DTDs, the cumulative 
fraction of explosions smoothly 
increases with the delay time, so that the fraction of {\it prompt} events 
depends 
on the definition of the timescale over which the events are considered as 
such.
As shown in Figs \ref{morph} and \ref{pritchet} these DTDs, with no strong
discontinuity, do account well for the observed trends of the SN Ia rates 
with the parent stellar population.

%mixed models: main features and comparison to observed correlations
% prediction about prob of getting a SD or DD explosion
% NO CONCLUSION ON THE SHAPE, BUT CONCLUSION ON KAIA
Motivated by recent observations which suggest that both SD and DD explosions 
may be realised in nature, I have constructed mixed models under two 
extreme assumptions: the \textit{Solomonic} option, in which the two channels
contribute the same fraction of events within a Hubble time, and the 
\textit{Segregated} mix, in which the SD and DD channels feed distinct ranges
of delay times. Since the SD model has difficulties to account for late epoch
explosions, I assumed that it provides a very \textit{prompt} component of the
DTD. This composition allows one to introduce a sharp discontinuity 
when the channel producing the events changes; the position in delay time of 
this discontinuity has been (arbitrarily) fixed at 0.1 Gyr to mimic the DTD
proposed by \citet{mannu06}. In the attempt of constraining the DTDs, either
single channel or mixed models, I have considered the observed correlations 
between (i) the SN Ia rate per unit luminosity and the parent galaxy type 
(GC09), and (ii) the SN Ia rate per unit mass and the 
specific SF rate of the parent galaxy  \citep{sulli06}.
Theoretical expectations have been computed with a population synthesis 
technique, adopting a variety of SF histories for the galaxies. To
construct the first of the two correlations, the galaxies' $(U-V)$ 
and $(B-K)$ colours
have been used to constrain the SF history; correlation (ii) turns out
largely independent of the adopted shape for the SF, as found by \citet{pri}.
The observed correlation between the SN Ia rate per unit luminosity and the
parent galaxy morphological type appears to require a substantial depletion 
of the DTD at 
late delay times. This holds especially for the rate measured in SNu's and 
derives from the drop of the rate going from S0 to E galaxies, which is not
detectable when the two classes are merged together. In fact, the colours of
S0s and Es indicate that the former are slightly 
younger, although both galaxy types are dominated by old stellar populations.
S0s and Es, then, probe the DTD at very late delay times, where, e.g., the 
SD model predicts a considerable drop. 
The steepening of the rate going from S0 to 
Es measured in GC09 supports DTDs with a steep decline at late delays, 
as the SD channel, or Solomonic mixtures. The DD channel and the Segregated 
mix could also meet this constraint, provided that the distribution of 
the separations is skewed at the short end.
%Therefore, the rates as a function of 
%morphological type as measured in GC09 support the SD channel or the
%DD channel with a distribution of separations skewed at the short end, as
%well as Solomonic and Segregated mixtures constructed with such single channel
%DTDs. 
This interpretation needs to be confirmed with larger galaxy samples and
more SN Ia events.

The correlation between the SN Ia rate and the galaxy type does not
provide a robust constraint on the shape of the DTD at short delay times,
because the SF history in the latest galaxies is not known with
sufficient accuracy, and, admittedly, the colours of Irregulars are not
well reproduced by the populations synthesis presented here. The 
theoretical rates are found consistent with the data for Irregulars and Sc 
galaxies even with the relatively low fraction of {\it prompt} events of 
the SD 
model, but a better understanding of the recent and past SF history in these
galaxies is needed to derive robust conclusions. 

The correlation between the
SN Ia rate per unit mass and the SSFR is compatible with the models for 
most of the explored DTDs, although the 
rates in the star bursting galaxies indicate that the DTD should not be
too skewed at the short delay times, thus disfavouring very steep DD CLOSE 
models and Segregated mixtures. 
%This applies to the average value of the
%DTD at $\tage \lesssim$ 0.5 Gyr, because of the \citet{sulli06} data binning.
Models computed with the SD DTD yield a very nice description of the 
correlation for the star forming galaxies; the rate in passive galaxies may 
also be well reproduced with this DTD if their inhabiting stellar populations
have a (small) age spread: e.g., one model galaxy  
having formed $\sim$ 80\% of its current mass within the first 1.7 Gyr, 
at an age of 9 Gyr has a SN Ia rate
\footnote{This estimate assumes $\kaia=10^{-2.8} \msun^{-1}$ and 37 \% a mass
return.}
of $\simeq 4.2\,10^{-14}$ (\msun yr)$^{-1}$ very close to what determined 
for the passive galaxies in \citet{sulli06} 
($\simeq 5\,10^{-14}$ (\msun yr)$^{-1}$). However, this estimate assumes that
all of the envelope of the secondary is accreted by the CO WD companion, 
which looks contrived.

\subsection[]{SN Ia productivity}

The fit of the observed relations yields a value for the productivity \kaia, 
i.e. 
the number of SN Ia's produced by a unit mass stellar population over the whole
range of delay times. In general \kaia\ depends on the IMF, DTD and
SF history assumed to model the galaxies, but the dependence on the DTD
is small for systems in which    
the SF rate varies little over the whole range of ages, and
actually vanishes for a constant SF rate (see Paper I). Therefore, in this paper,
the value of \kaia\ is derived by matching 
the level of the SN Ia rate in galaxies of intermediate type (Sa-Sb).
Assuming a bottom light IMF (Salpeter flattened below 0.5 \msun), 
from the rates in SNu's (SNuK's) 
in GC09 I find $\kaia \simeq 3.3 (2.8) \times 10^{-3} \msun^{-1}$, 
with a 50 \% total variation for all the DTDs plotted in Fig. \ref{morph}
(single and mixed channels). This value requires that $\simeq$ 8 \% 
of all stars with mass between 2.5 and 8 \msun\ produce a SN Ia event.
For the Segregated mixture, 30 \% of the progenitors are assumed to come
from systems with \msec\ between 5 and 8 \msun: this corresponds to a 
likelihood of the SN Ia event of $\sim 10\%$ from the more massive progenitors,
and of $\sim 4\%$ for stars with mass between 5 and 2.5 \msun.
 
It is instructive to compare the SN Ia to the core collapse (CC) SNe 
productivity
(\kcc) for the same IMF. The number of CC events per unit mass 
from a stellar generation depends on mass
limits of the progenitors; assuming an upper limit of 40 \msun, it amounts to  
6.5 (9) events every 1000 \msun\ for a lower limit of 10 (8) \msun, having
adopted a Salpeter flattened IMF.
Thus, each stellar generation is expected to produce $\sim$ 2-3 CC SNe per 
SN Ia; the events, though, occur at different delay times: all the CC  
explosions take place soon after the star formation episode, while the SN Ia 
events are diluted over the whole Hubble time, with only the {\it prompt} 
component exploding early. For the Segregated mixture considered here, a 
1000 \msun\ stellar populations provides $\simeq 1$ (\textit{prompt}) 
SN Ia within 0.1 Gyr from
birth; the same stellar population provides $\sim$ 7 CC SNe within 30 Myr from
birth. 
%Therefore, a galaxy sample probing the SNe produced by an 
%instantaneous burst of star formation within the first 0.1 Gyr of its life 
%should be characterised by a ratio of 7 CC to 1 Ia events.  
%
Therefore, a sample of elliptical galaxies supposed to have
experienced a starburst within the last 0.1 Gyr should yield 7 CC to 1
Ia {\it prompt} events; thus the excess of 7 SN Ia's in radio loud 
ellipticals noticed by 
\citet{dellavalle} should  be accompanied by 
$\sim 50$  CC events in the same galaxy sample, while none was observed. 
This observational result would require 
a detection efficiency of CC events $\sim 50$ times lower than that of 
Ia events, which is very hard to understand. The problem of 
the \textit{missing} CC SNe in radio loud Ellipticals becomes worse if
the CC productivity is higher, as implied, e.g.,  by a lower limit of 
8 \msun\ for their progenitors. 
It is worth noticing that although dust 
obscuration may preferentially hide CC SNe, since they explode soon after 
the onset of SF, in late type spirals 2.5 CC events
are detected every 1 SNIa \citep*{cbg07}, which shows that dust does not have
that dramatic effect at least in this kind of galaxies.

%Even accounting for the fact that 
%the lack of CC SNe in the Mannucci sample is hard to explain in this ...
%The ratio of
%observed events of the two types should be smaller than this, due to the 
%fact that CC SNe are more difficult to detect. Still, the sample of 
%radio loud ellipticals on which 
%higher difficulty of detecting CC SNe versus type Ia's. a short delay time from star formation should
%be characterized by many more CC SNe rather than SN Ia events. 
%If this amounts to $\sim$ 30 \% of the total events,
%the ratio between the number of CC and Ia SNe in a galaxy sample which  
%undergoes a star formation burst

The productivity \kaia\ can also be determined by fitting the rates 
from \citet{sulli06}. Matching the SN Ia rate in galaxies with 
SSFR = $10^{-10}$ (\msun yr)$^{-1}$   requires 
$\kaia \simeq 1.6 \times 10^{-3} \msun^{-1}$, which is $\sim 2$ times lower
than what estimated from the rates per unit luminosity. Notice that with 
this lower productivity, the problem of the \textit{missing} CC events in
radio loud Ellipticals worsens.
The discrepancy between the productivities estimated from the two kinds of
correlations may be due to the different astrophysical ingredients for
the stellar population models adopted here and in \citet{sulli06}: the IMF,
the SF history (i.e. age and metallicity distribution) of the galaxies, 
the stellar 
tracks to describe their light.
In fact, in order to obtain the observational SN Ia rates and SSFR, 
assumptions are necessary for all of these ingredients.
In addition, the estimates for \kaia\ derived here represent zero point 
determinations, 
rather than accurate fits to the data, and the true distance between the
best fitting \kaia\ values from the two correlations could be smaller than
what quoted above.
%Notice that also in this case the result depends on an assumed IMF, which
%must be specified in order to derive the observational rates per 
%unit mass. In addition, since the observed rates are normalized to the current
%mass in stars, an assumption on the IMF is necessary to apply the
%correction for the mass return to the model rates. In \citet{sulli06} galaxy 
%masses are determined from their SED with the P\'EGASE code and 
%\citet{kroupa} 
%IMF, which is similar to, but not exactly the same as adopted here to compute
%the mass return. Other factors may introduce a systematic difference  
%between the two \kaia\ estimates, including the tracks used for the population
%sysntehsis modelling (which impact on the mass to light ratios and on the
%SEDs), and the SF histories (i.e. the age and metallicity distributions) 
%of the galaxies. In addition, these values of \kaia\ do not follow from 
%accurate data fitting; rather they are equivalent to a zero point 
%determination.
All in all, the results on the productivity of SN Ia's from 
matching the observed rates are rather encouraging, though there seems to
be a systematic difference between the values derived from the two
data sets, with the rate per unit luminosity requiring a factor of $\sim$ 2 
higher productivity. Whether this is a robust conclusion or not can be
established only with complete
consistency of assumptions in modelling and observational estimates.

\begin{figure*}
\resizebox{\hsize}{!}{\includegraphics{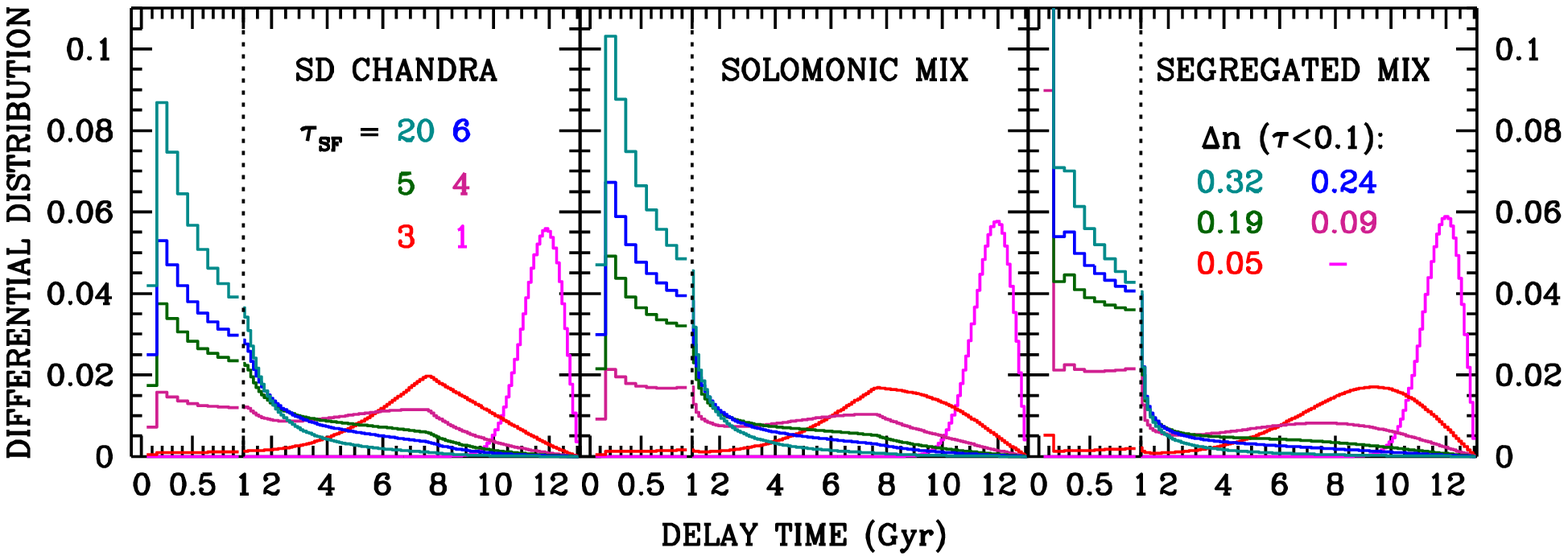}}
\resizebox{\hsize}{!}{\includegraphics{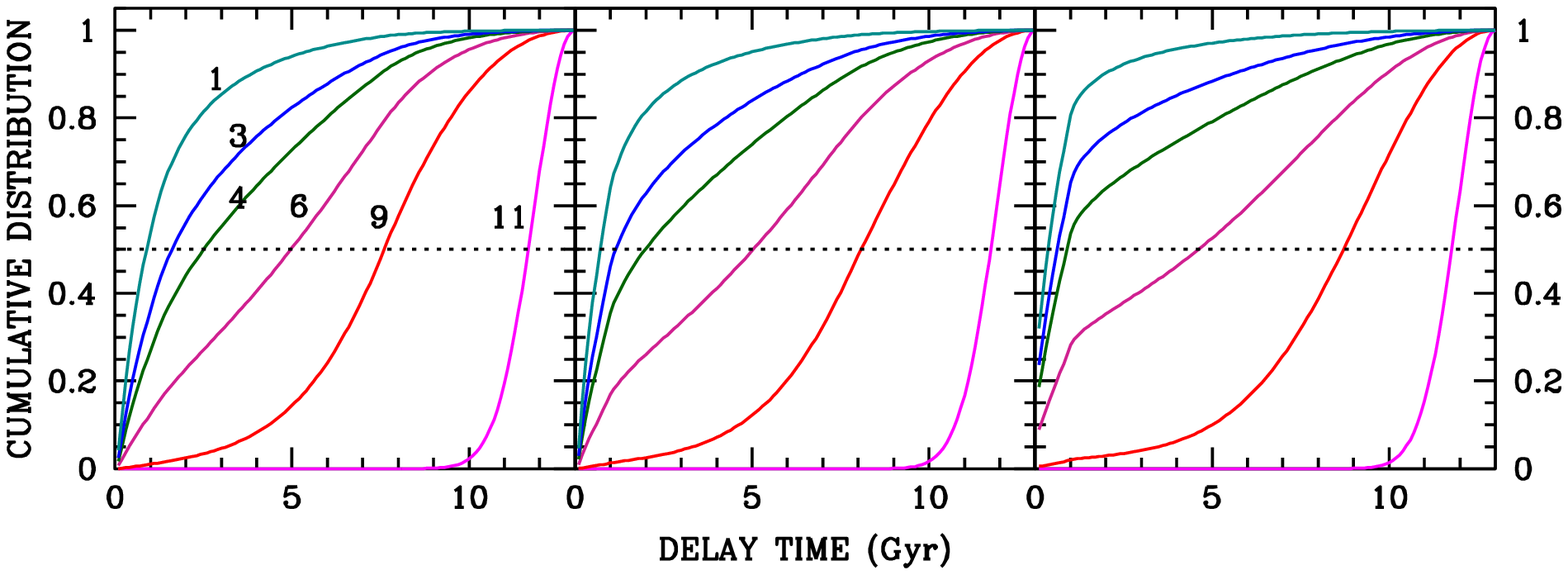}}
\caption{Differential (upper panels) and cumulative (lower panels) 
distributions of the delay times of events exploding in 13 Gyr old galaxies.
Different panels show the results for different DTDs, as labelled in the
upper figure; the mixed models assume a DD component of the CLOSE variety 
with \taunx=1 and \betag=-0.975 (DDC3). The colour encodes the SF history, 
given by Eq.\ref{eq_psi} with different values of $\tau_{SF}$, labelled in
the upper left panel. The values 20, 6, 5, 4, 3, and 1 Gyr produce
model galaxies with colours akin to Irregulars, Sc, Sb, Sa, S0 and E, 
respectively. For the Segregated mixture, the fraction of 
events with delay time shorter than 0.1 Gyr obtained with the different 
SF histories is labelled in the upper right panel.
The curves in the lower left panel are labelled with 
the (B luminosity weighted) average age of the galaxy model. 
Notice that the delay time scale in the upper panels has been expanded between
0 and 1 Gyr for better readability
of the figure.}
\label{taudist}
\end{figure*}

\subsection[]{Diversity of SN Ia events}

The astrophysical parameter responsible for the diversity of \nira\ production,
and then of SN Ia light curves, 
has not been identified yet. A number of possibilities have been proposed in 
the literature, some of which related to the progenitor, like metallicity,
age, ignition density, others more related to the physical phenomena taking
place during the explosion (\citealt{hoeko}; \citealt*{kasen}). 
From an empirical point of 
view, the most
conspicuous evidence concerns the correlation between the SN Ia light curve 
properties and the (average) 
age of the parent stellar population \citep{galla,howell}, metallicity
playing, in case, a secondary role. This suggests the possibility that the
diversity of SN Ia's is strongly related to their delay time. 

According to stellar evolution one  
expects a trend for the average mass and temperature of the WD progenitor with
delay time, but with a substantial spread. 
For the SD model, the average 
mass of the CO WD at the start of accretion first decreases, reaches a minimum,
but then increases again, since at the late delay times only massive 
CO WDs are able
to grow up to the Chandrassekhar limit. The range encompassed by the mass of
the CO WD is maximum at $\tage \sim$ 1 Gyr, but becomes narrower
at late delays. For the DD WIDE channel, the total mass of the successful 
DD system
is expected to vary between 1.4 and 2.4 \msun\ (almost) irrespective of the
delay time, since for each combination of \mpri\ and \msec,
small and wide separations should be realised, thereby providing events
with any gravitational wave radiation delay. For the DD CLOSE channel, instead,
we expect a trend of decreasing average \mdd\ as \tage\ increases, because the
more massive the progenitors, the smaller the separation of the DD system,
and then the shorter the time needed to merge.
The time between the formation of the CO WD and its final explosion is an
increasing function of the delay time. Thus,
for the DD CLOSE channel, the progenitors of events at short delay 
times are massive and hot (when accretion starts), while at long delay
times they are of low mass and cool. According 
to \citet{lesaffre} models these two
parameters could induce compensating trends on the ignition density, and 
the \nira\ mass synthesised in the 
explosion may not show big variations over the total range of delay times.
It is worth pointing out, however, that the models explore a range of 
relatively short 
cooling times (up to 1 Gyr), while the WD progenitors of SN Ia's in early
type galaxies have been cooling for many Gys. The outcome
of the explosion of CO WDs which start accreting when their structure is so
highly degenerate needs to be investigated. 

Regardless the key astrophysical variable which causes the diversity of
SN Ia light curves (whether ignition conditions, metallicity,
evolutionary channel leading to the explosion), if this variable 
scales systematically with the delay time one expects a trend of the
typical properties of SN Ia light
curve with the age distribution in the parent stellar population.
The observed correlation will depend on the SF history of the galaxies and on
the DTD: the younger the stellar populations
and/or the higher the fraction of short delay times, the higher the
likelihood that a detected SN Ia event has a short \tage. 

As an illustrative example, 
Fig.\ref{taudist} shows the distribution of the delay times of events 
occurring in galaxies with 13 Gyr of age, having experienced different SF 
histories. The differential distributions are computed as:

\begin{equation}
\Delta n = \frac{\int_\tau^{\tau+\Delta \tau} \psi(13-\tau)\, \fia(\tau)\, 
\rmn{d}\tau}{\int_0^{13} \psi(13-\tau)\, \fia(\tau)\, 
\rmn{d}\tau}
\end{equation} 

\acap
and represent the probability that a SN Ia detected in a galaxy of a certain 
type has a given delay time. 
%The differential distributions are normalized to 1 over the whole
%\tage\ range; thus they also represent the probability 
%that a detected SN Ia has a certain delay time.
It appears that in the late galaxy types most of the events cover a range of
delay times up to about 1-2 Gyr: as shown in Sect. 3 this is wide enough to
encompass the whole range of masses of the CO WDs. Only in the case 
of strongly star bursting galaxies the majority of events may come from
a truly \textit{prompt} component, i.e. the most massive and hottest CO WDs.
Even for the Segregated mixture in combination with the model for Irregular 
galaxies, more than 40 \% of the SN Ia's should have
a delay time in excess of 0.5 Gyr. In Sa galaxies all
the delay times have about the same probability of being sampled, and the 
distribution remains wide also in S0 galaxies, though mostly confined at
$\tage \gtrsim 5$ Gyr; in Ellipticals, instead, we expect
to sample only the latest delay times. Therefore, if diversity were related 
to the delay
time, SN Ia's in Es should be rather uniform, while in later galaxy types
we should detect a large variety of light curve properties, as 
suggested by the data. 
Notice that in the DD WIDE channel the mass
range of the successful DD systems is not likely to become narrower at late
delays: for this model, if diversity was connected to \mdd, a wide range 
of light curve properties should be observed also in the earliest galaxy types.

Uniformity of events in old galaxies is supported in \citet{howell}, but
%when considering the relation between the \nira\ mass and the average age of
%the host stellar population things become more puzzling. The 
it is difficult to understand quantitatively the result that
galaxies with a light averaged age in excess of 3 Gyr host 
only faint supernovae (technically with a low ejected \nira\ mass).
The curves plotted in the bottom left panel of Fig.\ref{taudist} are labelled
with the B-light averaged age of each galaxy model. Only Irregulars and Sc 
galaxies appear younger than 3 Gyr, while Sb's and Sa's have an average
age of several Gyrs. In these galaxies the distribution of the delay times
of their hosted SN Ia's is predicted rather broad, which would imply a 
broad range of photometric properties, irrespective of the 
specific DTD model. Therefore, it is difficult to reconcile the notion that
delay time is primarily responsible for diversity of SN Ia events with 
\citet{howell} result. 

% It appears that in
%late type galaxies the events are distributed over a relatively wide range 
%of delay times, with a relevant fraction at $\tage \gtrsim$ 1 G even for
%the model akin to Irregular galaxies in combination with the  segregated
%mixture. In other words, the probability that a SN Ia occurred in a
%late type galaxy comes from the \textit{prompt} component (\tage < 0.5 G) 
%is less then 50 \% in the most favourable case explored here 
%40 \% of the events have a delay time shorter than 0.5 Gyr. 

%expected divesity: variation of the range of MW,MDD,tcool; BUT WHAT MATTERS
% ARE THE DISTRIBUTION OF THESE PARAMETERS WITHIN THE LIMITS..
% more theoretical work on explosions, with inital structures corresponding to
% old and highly degenerate COWDs

\section{Conclusions}

According to expectations of stellar evolution in close binaries the DTD is
a continuous distribution, skewed at the short delay times, but with no real
justification for a separate {\it prompt} component. 
Mixed models may implement this 
concept, e.g. considering that the two channels (SD and DD) feed distinct 
ranges of delay times. However, there's no clear theoretical reason why 
it should be so, and
the observed correlations between the specific rates and the parent galaxy 
properties do not appear to require it.

Indeed, the observed correlations do not favour mixed over single 
channel models. The indication of the two channels being active comes from 
the data of individual explosions. More data are needed to establish whether 
the SD explosions are favoured at short delay times, i.e. in young stellar 
populations, which could support the Segregated mixture models presented here.

The correlation between the SN Ia rate per unit B-luminosity and the parent
galaxy type suggests a steep drop of the DTD at late epochs, consistent with 
the SD or DD models with a distribution of separations depleted of wide 
systems. The correlation between the SN Ia rate per unit mass and the specific 
SF rate is (broadly) consistent with this kind of models, and seems
to require a relatively small fraction of {\it prompt} events. However, in 
order to
constrain the fraction of short delay times in the DTD we need a more 
accurate description of the SF history in galaxies with high SSFR and of the 
latest morphological types.

The observed rates allow us to evaluate the SN Ia productivity \kaia, 
which appears to be in the
range $(1.6 - 3.3 )~10^{-3} \msun^{-1}$. Although it is 
reassuring that the different data sets yield values which 
agree within a factor of $\sim 2$,  
there may be a systematic difference between the 
productivity evaluated from the rates per unit luminosity and per unit SSFR.
On the other hand, the ingredients adopted for the galaxies'  modelling
to fit the two correlations are not exactly the same. Therefore the discrepancy
on the SN Ia productivity may just reflect a theoretical uncertainty. 

Concerning the diversity of SN Ia light curves, we expect a wide range for the 
mass of the CO WDs at the start of the accretion phase in most galaxy types;
only in the older stellar populations the properties of the progenitors should
be more uniform. For mixed models, in both the Solomonic and Segregated
flavours, late type galaxies should host
SD and DD events with comparable probability, while in early type galaxies
most, if not all, SN Ia should come from DD progenitors. Only in very young
stellar populations, with very high specific SF, the SD events are likely
to prevail. Current data on the correlation between the light curve 
properties and the host galaxy seem to yield contradictory indications,
possibly because of the use of different parameters to characterise the
SN Ia light curve and the properties of the host galaxy. 
Progress requires big samples of events, as well as more accurate knowledge of
the age distribution in SN Ia hosts; on the theoretical side more models
are also needed, in particular exploring the fate of an accreting CO WD 
which has been cooling for several Gyrs, hence 
starting from highly degenerate conditions.

A broader diversity of progenitor's properties is expected in local
SN Ia samples compared to high redshift ones, since the former encompass all
possible delay times, while the latter necessarily lack events with delay times
longer than the corresponding age of the universe. Nevertheless, a fair 
fraction of 
explosions in nearby star forming galaxies (spirals and irregulars) should
have a short delay time, e.g. shorter than 1 Gyr. Therefore the empirical 
\textit{standardisation} of the light curve obtained from the SN Ia's in
nearby star forming galaxies should be applicable also to high redshift 
events, thus supporting SN Ia's as viable standard candles all the way to 
cosmological distances.
 
%Due to the presence of the long delay times component, diversity of the 
%progenitors' properties is likely more 
%pronounced in local SN Ia samples with respect to high redshift ones.
%At the same time, though, a fair fraction
%of events in nearby Spiral and Irregular galaxies should be \textit{young},
%i.e. with a delay time shorter than 1 Gyr. This time scale is long enough to
%encompass the full range of CO WD progenitor masses.
%Therefore, the \textit{standardization} of
%the light curve obtained from SN Ia's in star forming nearby galaxies
%should be applicable also to high redshift events.
%\textbf {DISCUSSION: table prompt/tardy, diversity as lower and upper limits,
%but distributions...whatever, tauni lower connected to evol lifetime of
%maximum \msec  WAITING FOR LOSS BY FILIPPENKO}

\section*{Acknowledgments}
I am indebted to Alvio Renzini for helpful discussions and comments, and for 
careful reading of the manuscript. I also thank Enrico Cappellaro for 
discussions and for checking on the discrepancy between the
SN Ia rate in early type galaxies determined by \citet{mannu05} and by GC09.
Finally, I thank the anonymous referee for productive comments.

\label{lastpage}


\begin{thebibliography}{99}
\bibitem[\protect\citeauthoryear{Abt}{1983}]{abt} Abt H.A., 1983, ARA\&A, 21, 343
\bibitem[\protect\citeauthoryear{Altavilla et al.}{2004}]{altav} Altavilla G. et al., 2004, MNRAS, 349, 1344
\bibitem[\protect\citeauthoryear{Arnett}{1982}]{arnett} Arnett D., 1982, ApJ, 253, 785
\bibitem[\protect\citeauthoryear{Belczynski, Bulik \& Ruiter}{Belczynski et al.}{2005}]{belczy05} Belczynski K., Bulik T., Ruiter A.J., 2005, ApJ, 629, 915
\bibitem[\protect\citeauthoryear{Bessell}{1990}]{bessell} Bessell M.S., 1990, Publ. Astron. Soc. Pac., 102, 1181
\bibitem[\protect\citeauthoryear{Blanc \& Greggio}{2008}]{blanc} Blanc G., Greggio L., 2008, New Astron., 13, 606
\bibitem[\protect\citeauthoryear{Cappellaro, Evans \& Turatto}{Cappellaro et al.}{1999}]{cet99} Cappellaro E., Evans R., Turatto M., 1999, A\&A, 351, 459 
\bibitem[\protect\citeauthoryear{Cappellaro, Botticella \& Greggio}{Cappellaro et al.}{2007}]{cbg07} Cappellaro E., Botticella M.T., Greggio L., 2007, in Immler S., Weiler K., McCray R., eds, AIP Conf. Proc. 937, Supernova 1987A:20 Years After: Supernovae and Gamma-Ray Bursters. Am. Inst. Phys., New York, p. 198 
\bibitem[\protect\citeauthoryear{Ciotti et al.}{1991}]{cio91} Ciotti L., D'Ercole A., Renzini A., Pellegrini S., 1991, ApJ, 376, 380 
\bibitem[\protect\citeauthoryear{Della Valle et al.}{2005}]{dellavalle} Della Valle M., Panagia N., Padovani P., Cappellaro E., Mannucci F., Turatto M., 2005, ApJ, 629,750
\bibitem[\protect\citeauthoryear{de Vaucouleurs, de Vaucouleurs \& Corwin}{de Vaucouleurs et al.} {1991}]{devau} de Vaucouleurs G., de Vaucouleurs A., Corwin H.G., Buta R.J., Paturel G., Fouque P. , 1991. Third Reference Catalogue of Bright Galaxies. Springer-Verlag, New York
\bibitem[\protect\citeauthoryear{Di Stefano et al.}{2009}]{distefano} Di Stefano R., Primini F.A., Liu J., Kong A., Patel B., 2009, preprint (arXiv:0909.2046) 
\bibitem[\protect\citeauthoryear{Ellis et al.}{2008}]{ellis} Ellis, R.S. et al., 2008, ApJ, 674, 51 
\bibitem[\protect\citeauthoryear{Fioc \& Rocca-Volmerange}{1997}]{rocca97} Fioc M., Rocca-Volmerange B., 1997, A\&A, 326, 950
\bibitem[\protect\citeauthoryear{Gallagher et al.}{2008}]{galla} Gallagher J.S., Garnavich P.M., Caldwell N., Kirshner R.P., Jha S.W., Li W., Ganeshalingam M., Filippenko A,.V., 2008, ApJ, 685, 752 
%\bibitem[\protect\citeauthoryear{Gallagher et al.}{2008}]{galla} Gallagher J.S., Garnavich P.M., Caldwell N., Kirshner R.P., Jha S.W., 2008, ApJ 685, 752 
\bibitem[\protect\citeauthoryear{Gavazzi et al.}{2002}]{gavazzi} Gavazzi G., Bonfanti C., Sanvito G., Boselli A., Scodeggio M., 2002, ApJ, 576, 135
\bibitem[\protect\citeauthoryear{Greggio}{1996}]{g96} Greggio L., 1996, in Kunth D., Guiderdoni B., Heydari-Malayeri M., Thuan T.X., eds. The Interplay Between Massive Star Formation, the ISM, and Galaxy Evolution. Edition Fronti\'eres, Gif-sur-Yvette, p. 98 
\bibitem[\protect\citeauthoryear{Greggio}{2005}]{g05} Greggio L., 2005, A\&A, 441, 1055 (Paper I)
\bibitem[\protect\citeauthoryear{Greggio \& Cappellaro}{2009}]{gc09} Greggio L., Cappellaro E., 2009, in Antonelli L.A., Brocato E., Limongi M., Menci N., Raimondo G., Tornamb\'e A., eds, AIP Conf. Proc. 1111. Probing Stellar Populations Out to the Distant Universe. Am. Inst. Phys., New York, p.477 (GC09)
%Probing Stellar Populations out to the Distant Universe, L.A. Antonelli, E. Brocato, M. Limongi, M. Menci, G. Raimondo and A. Tornamb\'e eds   
\bibitem[\protect\citeauthoryear{Greggio \& Renzini}{1983}]{gr83} Greggio L., Renzini A., 1983, A\&A, 118, 217  
\bibitem[\protect\citeauthoryear{Greggio, Renzini \& Daddi}{Greggio et al.}{2008}]{grd} Greggio L., Renzini A., Daddi E., 2008, MNRAS, 388, 829 
%\bibitem[\protect\citeauthoryear{Mannucci et al.}{2005}]{mannu05} Mannucci F., Della Valle M., Panagia N., Cappellaro E., Cresci G., Maiolino R., Petrosian A., Turatto M., 2005, A\&A, 433, 807
\bibitem[\protect\citeauthoryear{Hachisu, Kato \& Nomoto}{Hachisu et al.}{1996}]{hkn96} Hachisu I., Kato M., Nomoto K., 1996, ApJ, 470, L100 
\bibitem[\protect\citeauthoryear{Hachisu, Kato \& Luna}{Hachisu et al.}{2007}]{hachi} Hachisu I., Kato M., Luna G.J.M., 2007, ApJ, 659, L153 
\bibitem[\protect\citeauthoryear{Hachisu, Kato \& Nomoto}{Hachisu et al.}{2008}]{hkn08} Hachisu I., Kato M., Nomoto K., 2008, ApJ, 683, L127 
\bibitem[\protect\citeauthoryear{Han \& Podsiadlowski}{2004}]{han04} Han Z., Podsiadlowski Ph., 2004, MNRAS, 350, 1301
\bibitem[\protect\citeauthoryear{Hicken et al.}{2009}]{hicken} Hicken M. et al.
, 2009, ApJ, 700, 331
\bibitem[\protect\citeauthoryear{Hillebrandt \& Niemeyer}{2000}]{hille} Hillebrandt W., Niemeyer J.C., 2000, ARA\&A, 38, 191
\bibitem[\protect\citeauthoryear{Hoeflich \& Khokhlov}{1996}]{hoeko} Hoeflich P., Khokhlov A., 1996, ApJ, 457, 500
\bibitem[\protect\citeauthoryear{Hoeflich et al.}{1996}]{hoefli} Hoeflich P., Khokhlov A., Wheeler J.C., Phillips M.M., Suntzeff N.B., Hamuy M., 1996, ApJ, 472, L81
\bibitem[\protect\citeauthoryear{Howell et al.}{2006}]{howell06} Howell D.A. et al., 2006, Nature, 443, 308
\bibitem[\protect\citeauthoryear{Howell et al.}{2009}]{howell} Howell D.A. et al., 2009, ApJ, 691, 661 
\bibitem[\protect\citeauthoryear{Iben}{1991}]{iben91} Iben I.Jr, 1991, ApJS, 76, 555
\bibitem[\protect\citeauthoryear{Iben \& Tutukov}{1987}]{it87} Iben I.Jr, Tutukov A.V., 1987, ApJ, 313, 727
\bibitem[\protect\citeauthoryear{Iben \& Tutukov}{1994}]{it94} Iben I.Jr, Tutukov A.V., 1994, ApJ, 431, 264
\bibitem[\protect\citeauthoryear{Lesaffre et al.}{2006}]{lesaffre} Lesaffre P., Han Z., Tout C.A., Podsiadlowski Ph., Martin R.G., 2006, MNRAS, 368, 187
\bibitem[\protect\citeauthoryear{Kalirai et al.}{2008}]{kalirai} Kalirai J.S., Hansen B.M.S., Kelson D.D., Reitzel D.B., Rich R.M., Richer H.B., 2008, ApJ, 676, 594
\bibitem[\protect\citeauthoryear{Kasen}{2010}]{kasen10} Kasen D., 2010, ApJ, 708, 1025
\bibitem[\protect\citeauthoryear{Kasen, Roepke \& Woosley}{Kasen et al.}{2009}]{kasen} Kasen D., Roepke F.K., Woosley S.E., 2009, Nature, 460, 869
\bibitem[\protect\citeauthoryear{Kenyon et al.}{1993}]{kenyon} Kenyon S.J., Livio M., Mikolajewska J., Tout C.A., 1993, ApJ, 407, L81
\bibitem[\protect\citeauthoryear{Kouwenhoven et al.}{2007}]{kouwen} Kouwenhoven M.B.N., Brown A.G.A., Portegies Zwart S.F., Kaper L., 2007, A\&A, 474, 77
\bibitem[\protect\citeauthoryear{Kroupa}{2001}]{kroupa} Kroupa P., 2001, MNRAS, 322, 231
\bibitem[\protect\citeauthoryear{Ma\'iz Appell\'aniz}{2006}]{maiz} Ma\'iz Appell\'aniz J., 2006, AJ, 131, 1184
\bibitem[\protect\citeauthoryear{Mannucci et al.}{2005}]{mannu05} Mannucci F., Della Valle M., Panagia N., Cappellaro E., Cresci G., Maiolino R., Petrosian A., Turatto M., 2005, A\&A, 433, 807
\bibitem[\protect\citeauthoryear{Mannucci, Della Valle \& Panagia}{Mannucci et al.}{2006}]{mannu06}Mannucci F., Della Valle M., Panagia N., 2006, MNRAS, 370, 773 
\bibitem[\protect\citeauthoryear{Matteucci \& Greggio}{1986}]{mg86} Matteucci M.F., Greggio L., 1986, A\&A, 154, 279
\bibitem[\protect\citeauthoryear{Matteucci et al.}{2006}]{chicchi06} Matteucci M.F., Panagia N., Pipino A., Mannucci F., Recchi S., Della Valle M., 2006, MNRAS, 372, 265
\bibitem[\protect\citeauthoryear{Marigo et al.}{2008}]{marigo08} Marigo P., Girardi L., Bressan A., Groenewegen M.A.T., Silva L., Granato G.L., 2008, A\&A, 482, 883
\bibitem[\protect\citeauthoryear{Mazzali et al.}{2001}]{mazzali} Mazzali P.A., Nomoto K., Cappellaro E., Nakamura T., Umeda H., Iwamoto K., 2001, ApJ, 547, 988
\bibitem[\protect\citeauthoryear{Meng, Chen \& Han}{Meng et al.}{2009}]{meng} Meng X., Chen X., Han Z., 2009, MNRAS, 395, 2103
\bibitem[\protect\citeauthoryear{Mereghetti et al.}{2009}]{mereghe} Mereghetti S., Tiengo A., Esposito P., La Palombara N., Israel G.L., Stella L., 2009, Sci, 325, p. 1222
\bibitem[\protect\citeauthoryear{Munari \& Renzini}{1992}]{munari} Munari U., Renzini A., 1992, ApJ, 397, L87
\bibitem[\protect\citeauthoryear{Napiwotzki et al.}{2005}]{napi} Napiwotzki R.,
et al., 2005,in Koester D., Moehler S, eds, ASP Conf. Series 334. 14th European Workshop on White Dwarfs. Astron. Soc. Pac., San Francisco, p.375
\bibitem[\protect\citeauthoryear{Neill et al.}{2009}]{neill} Neill J.D. et al., 2009, ApJ, 707, 1449
\bibitem[\protect\citeauthoryear{Nelemans et al.}{2000}]{nele00} Nelemans G., Verbunt F., Yungelson L.R., Portegies-Zwart S.F., 2000, A\&A, 360, 1011 
\bibitem[\protect\citeauthoryear{Nobili \& Goobar}{2008}]{nobili} Nobili S.,
Goobar A., 2008, A\&A, 487, 19 
\bibitem[\protect\citeauthoryear{Nomoto \& Iben}{1985}]{ken85} Nomoto K.,
Iben I.Jr, 1985, ApJ, 297, 531 
\bibitem[\protect\citeauthoryear{Nomoto et al.}{2009}]{ken09} Nomoto K., Kamiya Y., Nakasato N., Hachisu I., Kato M., 2009, in Antonelli L.A., Brocato E., Limongi M., Menci N., Raimondo G., Tornamb\'e A., eds, AIP Conf. Proc. 1111. Probing Stellar Populations Out to the Distant Universe. Am. Inst. Phys., New York, p.267
\bibitem[\protect\citeauthoryear{Nugent et al.}{1997}]{nugent} Nugent P., Baron E., Branch D., Fisher A., Hauschildt P.H., 1997, ApJ, 485, 812
%Probing Stellar Populations out to the Distant Universe, L.A. Antonelli, E. Brocato, M. Limongi, M. Menci, G. Raimondo and A. Tornamb\'e eds   
\bibitem[\protect\citeauthoryear{Oemler \& Tinsley}{1979}]{oemler} Oemler A., Tinsley B.M. , 1979, AJ, 84, 985
\bibitem[\protect\citeauthoryear{Parthasarathy et al.}{2007}]{partha} Parthasarathy M., Branch D., Jeffery D.J., Baron E., 2007, New Astron. Rev., 51, 524
\bibitem[\protect\citeauthoryear{Patat et al.}{2007}]{patat} Patat F. et al. , 1997, Sci., 317, 924
\bibitem[\protect\citeauthoryear{Perlmutter et al.}{1997}]{perl97} Perlmutter S. et al. , 1997, ApJ, 483, 565 
\bibitem[\protect\citeauthoryear{Perlmutter et al.}{1999}]{perl} Perlmutter S. et al. , 1999, ApJ, 517, 565 
\bibitem[\protect\citeauthoryear{Phillips}{1993}]{phillips} Phillips M.M., 1993, ApJ, 413, L105 
\bibitem[\protect\citeauthoryear{Piersanti et al.}{2009}]{pier09} Piersanti L., Tornamb\'e A., Straniero O., Dom\'inguez I., 2009, in Antonelli L.A., Brocato E., Limongi M., Menci N., Raimondo G., Tornamb\'e A., eds, AIP Conf. Proc. 1111. Probing Stellar Populations Out to the Distant Universe. Am. Inst. Phys., New York, p.259
%Probing Stellar Populations out to the Distant Universe, L.A. Antonelli, E. Brocato, M. Limongi, M. Menci, G. Raimondo and A. Tornamb\'e eds   
\bibitem[\protect\citeauthoryear{Pinsonneault \& Stanek}{2006}]{pinso}
  Pinsonneault M.H., Stanek K.Z., 2006, ApJ, 639, L67 
\bibitem[\protect\citeauthoryear{Pritchet, Howell \& Sullivan}{Pritchet et al.}{2008}]{pri} Pritchet C.J., Howell D.A., Sullivan M., 2006, ApJ, 683, L25 
\bibitem[\protect\citeauthoryear{Riess et al.}{1998}]{riess} Riess A.G. et al. , 1998, AJ, 116, 1009 
\bibitem[\protect\citeauthoryear{Renzini}{2004}]{alvio04} Renzini A., 2004, in 
Mulchaey J.S., Dressler A., Oemler A., eds. Clusters of Galaxies: Probes of Cosmological Structure and Galaxy Evolution.
Cambridge Univ. Press, Cambridge, p. 260 
\bibitem[\protect\citeauthoryear{Renzini}{2006}]{alvioara} Renzini A., 2006, ARA\&A, 44, 141 
\bibitem[\protect\citeauthoryear{Ruiter, Belczynski \& Fryer}{Ruiter et al.}{2009}]{belczy09} Ruiter A.J., Belczynski K., Fryer C., 2009, ApJ, 699, 2026
%Proceedings of the conference held 15-19 June, 2004 in Padua, Italy. Edited by M. Turatto, S. Benetti, L. Zampieri, and W. Shea. San Francisco: Astronomical Society o
\bibitem[\protect\citeauthoryear{Salaris et al.}{2000}]{salaris} Salaris M., Garc\'ia-Berro E., Hernandez M., Isern J., Saumon D., 2000, ApJ, 544, 1036 
\bibitem[\protect\citeauthoryear{Simon et al.}{2009}]{simon} Simon J.D. et al., 2009, ApJ, 702, 1157 
\bibitem[\protect\citeauthoryear{Sullivan et al.}{2005}]{sulli05} Sullivan M. et al. , 2005, in Turatto M., Benetti S., Zampieri L., Shea W., eds, ASP Conf. Series 342. 1604-2004: Supernovae as Cosmological Lighthouses. Astron. Soc. Pac., San Francisco, p.466.
\bibitem[\protect\citeauthoryear{Sullivan et al.}{2006}]{sulli06} Sullivan M. et al. , 2006, ApJ, 648, 868 
\bibitem[\protect\citeauthoryear{Tammann}{1978}]{tammann}Tammann G.A., 1977, Mem. Soc. Astron. Ital., 49, 315 
\bibitem[\protect\citeauthoryear{Timmes, Brown \& Truran}{Timmes et al.}{2003}]{timmes}Timmes F.X., Brown E.F., Truran J.W., 2003, ApJ, 590, L83 
\bibitem[\protect\citeauthoryear{Trimble}{2008}]{trimble} Trimble V., 2008, The  Observatory, 128, 286
\bibitem[\protect\citeauthoryear{van den Heuvel et al.}{1992}]{vandenh} van den Heuvel E.P.J., Bhattacharya D., Nomoto K., Rappaport S., 1992, A\&A, 262, 97
\bibitem[\protect\citeauthoryear{Yamanaka et al.}{2009}]{yamana} Yamanaka M., et al., 2009, ApJ, 707, L118
\bibitem[\protect\citeauthoryear{Yoon, Podsiadlowski \& Rosswog}{Yoon et al.}{2007}]{yoon} Yoon S.-C., Podsiadlowski Ph., Rosswog S., 2007, MNRAS, 380, 933
\bibitem[\protect\citeauthoryear{Yungelson}{2005}]{yungelson} Yungelson L.R., 2005,in Burderi L., Antonelli L.A., D'Antona F., Di Salvo T., Israel G.L., Piersanti L., Tornamb\'e A., Straniero O., eds,  AIP Conf. Proc. 797. Interacting Binaries: Accretion, Evolution and Outcomes. Am. Inst. Phys., New York, p.1
\bibitem[\protect\citeauthoryear{Yungelson \& Livio}{2000}]{yungli} Yungelson L.R., Livio M., 2000, ApJ, 528, 108
\bibitem[\protect\citeauthoryear{Wang et al.}{2009}]{wang} Wang B., Meng X., Chen X., Han Z., 2009, MNRAS, 395, 847
\bibitem[\protect\citeauthoryear{Webbink}{1984}]{webbink} Webbink R.F., 1984, ApJ, 227, 355
\bibitem[\protect\citeauthoryear{Williams, Bolte \& Koester}{Williams et al.}{2004}]{willi} Williams K.A., Bolte M., Koester D., 2004, ApJ, 615, L49
\bibitem[\protect\citeauthoryear{Woosley \& Weaver}{1994}]{stan94} Woosley S.E., Weaver T.A., 1994, ApJ, 423, 371
\end{thebibliography}
\end{document}